\documentclass[natbib]{svjour3}    
\smartqed  
\usepackage{graphicx}

\usepackage{amsmath,amsbsy}
\usepackage{amssymb}
\usepackage{color}

\newcommand{\degr}{{^{\rm o}}}

\newcommand{\pasj}{PASJ}

\newcommand{\pasp}{PASP}
\newcommand{\apjs}{ApJS}
\newcommand{\apj}{ApJ}
\newcommand{\apjl}{ApJL}

\newcommand{\aap}{A\&A}

\newcommand{\aj}{AJ}
\newcommand{\mnras}{MNRAS}
\newcommand{\nat}{Nature}

\newcommand{\araa}{ARA\&A}
\newcommand{\aapr}{A\&ARv}          
          
\newcommand{\ssr}{Space Sci. Rev.}

\newcommand{\msun}{\mbox{M}_{\odot}}       
\newcommand {\beq}{\begin {eqnarray}}
\newcommand {\eeq}{\end {eqnarray}}

\begin{document}

\title{Modelling spectral and timing properties of  accreting black holes: 
the hybrid hot flow paradigm}


\titlerunning{Hybrid hot flow paradigm}        

\author{Juri  Poutanen         \and
        Alexandra Veledina 
}

\authorrunning{Juri  Poutanen         \and
        Alexandra Veledina} 

\institute{J. Poutanen, A. Veledina \at
Astronomy Division, Department of Physics, PO Box 3000, FI-90014 University of Oulu, Finland\\
\email{juri.poutanen@gmail.com, alexandra.veledina@oulu.fi}           
}

\date{Received: date / Accepted: date}

\maketitle

\begin{abstract}
The general picture that emerged by the end of 1990s from a large set 
of optical and X-ray, spectral and timing data was that 
the X-rays are produced in the innermost hot part of the accretion flow, 
while the optical/infrared (OIR) emission is mainly produced by the irradiated outer thin accretion disc. 
Recent multiwavelength observations of Galactic black hole transients 
show that the situation is not so simple. 
Fast variability in the OIR band, OIR excesses above the thermal emission  and 
a complicated interplay between the X-ray and the OIR light curves 
imply that the OIR emitting region is much more compact. 
One of the popular hypotheses is that the jet contributes to the OIR emission 
and even is responsible for the bulk of the X-rays. 
However, this scenario is largely ad hoc and is in contradiction with many previously established facts. 
Alternatively, the hot accretion flow,  
known to be consistent  with the  X-ray spectral and timing data, is also a viable candidate to produce the OIR radiation. 
The hot-flow scenario naturally explains the power-law like OIR spectra, fast OIR variability 
and its complex relation to the X-rays if the hot flow contains non-thermal electrons 
(even in energetically negligible quantities), 
which  are required by the presence of the MeV tail in Cyg~X-1. 
The  presence of non-thermal electrons also lowers the equilibrium electron temperature 
in the hot flow model to $\lesssim100$ keV, making it more consistent with observations. 
Here we argue that any viable model should simultaneously explain a large set of spectral and timing data
and show that the hybrid (thermal/non-thermal) hot flow model satisfies most of the constraints. 
\end{abstract}

\section{Introduction}
\label{sec:intro}

Models for accretion onto a black hole (BH) have been discussed now for more than 40 years.
During the last 10--15 years we have seen a dramatic increase in 
the amount of information on the BH X-ray binaries (BHBs). 
Spectral details (iron lines and Compton reflection), 
spectral transitions, and variability on various time scales has been studied 
in unprecedented details with the new generation X-ray telescopes such as 
{\it Rossi X-ray Timing Explorer (RXTE)} and {\it XMM-Newton}. 
Excellent recent reviews are devoted to these advances \citep{ZG04,RM06,DGK07,Done10,Gilfanov10}.
 
In addition to the X-ray data, we have seen an explosion of  
information coming from other wavelengths: radio, sub-mm, 
infrared, optical, UV, MeV and nowadays even from the GeV range. 
What is even more spectacular is that the properties of the 
BHs at these other wavebands are correlated with the X-ray flux and X-ray states. 
Among the most impressive achievements we find the discovery 
of correlated fast variability in the optical/infrared (OIR) band and in the X-rays 
\citep{Kanbach01,DGS08,GMD08} with some hints actually coming already 30 years ago \citep{Motch83}.
This got theoreticians to scratch their heads and invent new models that 
often were in disagreement with previously established theories and contradicted many other available data. 

Here we discuss some of the recent discoveries. We would like to note  
that  the time for theoretical (phenomenological) models 
based purely on spectral properties are long gone.  
In order to be considered seriously, 
any model has to address many observed facts together. 

This review consists of two parts. In the first one, 
we discuss the most recent reincarnation of the hot flow model, which 
now also considers the role of the non-thermal particles. 
In the second part, we discuss recent observational advances. 
We review the spectral data in various energy bands concentrating on the 
X-rays and the OIR. Then we discuss the observed temporal properties and correlated variability in different 
energy bands, as well as more complicated  temporal-spectral  statistics such as Fourier resolved spectra. 
In this review we will concentrate on the hard state and 
interpret the observations in terms of the hot flow model.

\section{Hot flow models} 
 
\subsection{Comptonization models for the X-ray emission} 
 
\begin{figure}
\centerline{\includegraphics[width=12cm]{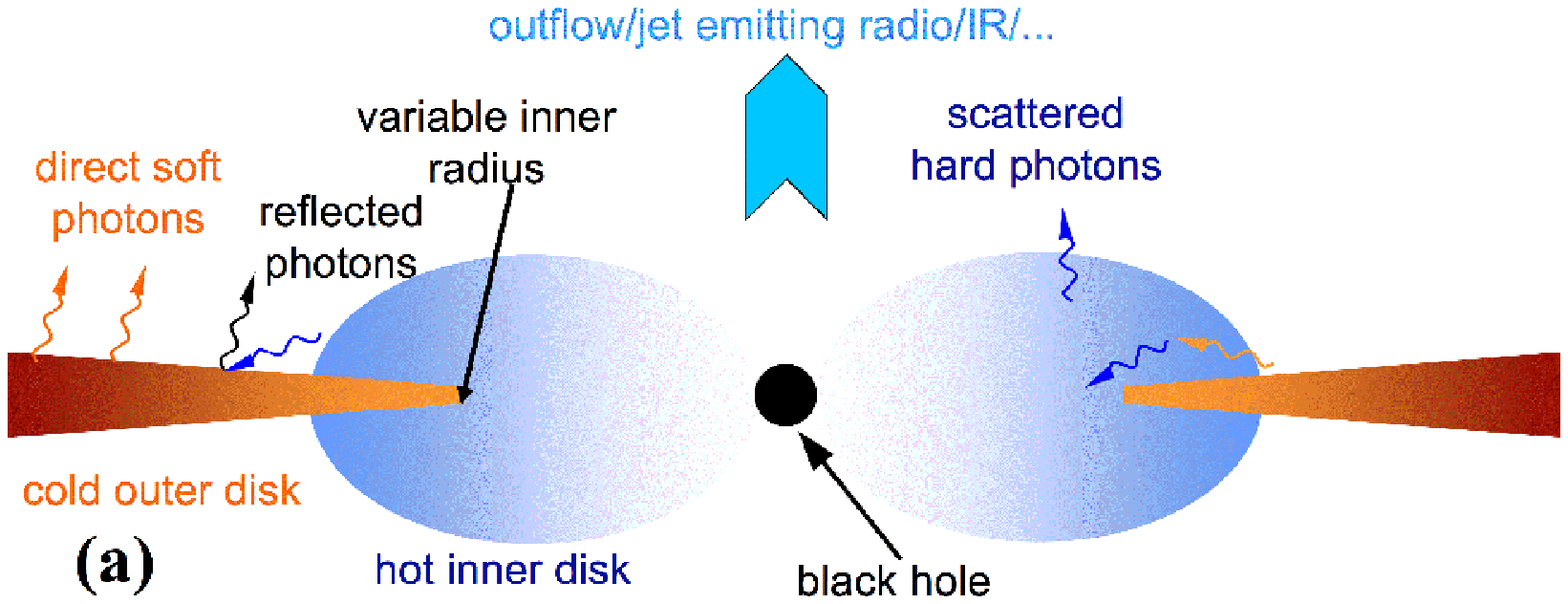}} 
\vskip 0.7cm
\centerline{\includegraphics[width=12cm]{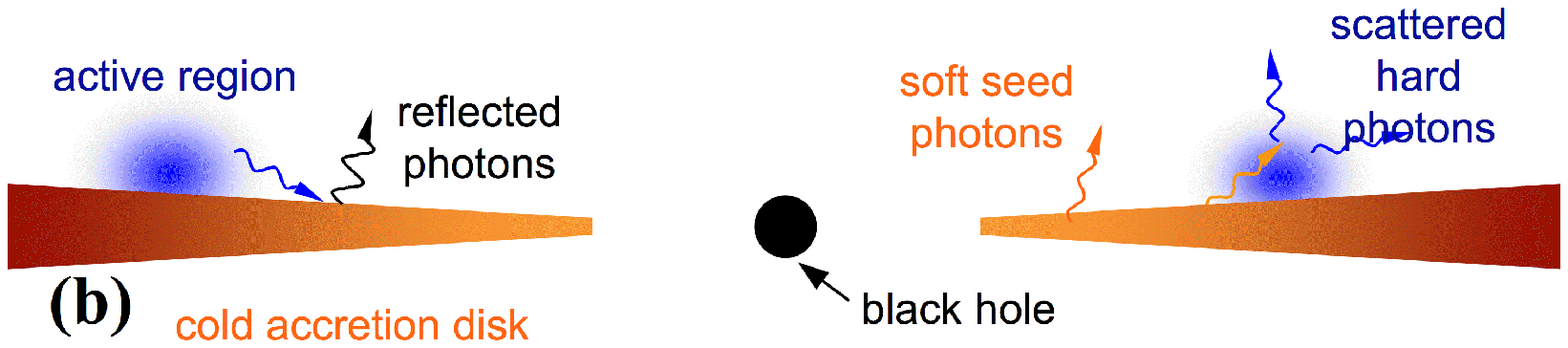}} 
\caption{(a) A schematic representation of the likely geometry in the hard state,
consisting of a hot inner accretion flow surrounded by
optically-thick cold accretion disc. The hot flow constitutes the base
of the jet (with the counter-jet omitted from the figure for
clarity). The disc is truncated far away from the minimum stable
orbit, but it may overlap with the hot flow. The soft photons emitted
by the disc  (and possibly internally produced synchrotron photons) 
are Compton upscattered in the hot flow, and emission
from the hot flow is partly Compton-reflected from the disc. (b)
The likely geometry in the soft state consisting of flares/active
regions above an optically-thick accretion disc extending close to
the minimum stable orbit. The soft photons emitted by the disc are
Compton upscattered in the flares  
by non-thermal electrons producing power-law spectra extending to $\gamma$-rays. 
Emission from the flares is partly Compton-reflected from the disc.   From \citet{ZG04}. 
 }
\label{fig:geom}
\end{figure}

At high accretion rates exceeding typically 10\% of the Eddington value, BHs are in the ``soft state'' and 
have thermally looking spectra peaking in the standard 2--10 keV X-ray band, 
which are consistent with the thin $\alpha$-disc model \citep{SS73,NT73}. 
These thermally-dominated spectra presumably depend only on the accretion rate, the BH mass and spin and the inclination. 
They potentially can  be used  to determine the BH spin if the distance to the source and e.g. the BH mass and inclination 
are known \citep[see][]{MCNS13SSRv}. 
However, often strong power-law tails are seen (see Fig.~\ref{fig:tail_cygx1}).
This tails are interpreted as a signature of non-thermal ``corona'' atop of the standard Shakura-Sunyaev disc (Fig.~\ref{fig:geom}b).

At the lower accretion rate, BHs are often found in the ``hard state'' and their spectra do not  even 
remotely look thermal, but are close to the power-law in the  X-ray band 
with a sharp cutoff at about 100 keV \citep[e.g.][]{G97,ZPM98}. 
These are well described  by the thermal Comptonization model \citep[see ][ for reviews]{ZJP97,P98}.  
The nature of the hard state emission and origin of the hot electrons has been already discussed in the 1970s 
\citep{SLE76,Ichimaru77} and is commonly associated with either a hot 
inner flow close to the BH \citep{Esin97,Esin98,PKR97,NMQ98,yuan_aaz04} or a corona 
above the accretion disc \citep[][see Fig.~\ref{fig:geom}a  for a possible 
geometry]{GRV79,HM93,HMG94,SPS95,PS96,B99PE}.

Knowing the slope of the hard state spectra (with photon index $\Gamma=1.6$--1.8)
we can easily estimate the ratio of the total emitted power $L$ to 
the soft seed photon luminosity $L_{\rm s}$ entering Comptonization region, 
i.e. the amplification factor $A=L/L_{\rm s}$. 
\citet{B99ASP} found an approximate relation between $\Gamma$ and $A$ for the  Comptonized  spectra: 
\begin{equation}\label{eq:gamma_delta}
\Gamma= \frac{7}{3} (A-1)^{-\delta}, 
\end{equation}
where $\delta=1/6$ for BHBs and  $\delta=1/10$ for AGNs and 
the typical seed photon temperatures of  0.2 keV for BHBs and 5 eV for AGNs were assumed.  
If indeed the disc photons are being Comptonized, we get  $A\approx 10$ for BHBs in their hard state. 
This fact puts serious constraints on the geometry of the emission region \citep{P98,B99ASP}
and  immediately rules out simple slab-corona models which predict much 
smaller amplification  $A\lesssim 2$ and softer  spectra because 
of the efficient X-ray reprocessing in the cold disc \citep{SPS95}. Assuming that coronal plasma has 
a mildly relativistic velocity  away from the cold disc \citep{B99PE,MBP01} 
one can in principle reconcile that model with the observed slopes as well as with the   
correlated changes of  the spectral hardness and the amount of Compton reflection from the disc, but 
still one would have troubles explaining their correlations  with 
the iron line width and characteristic variability frequencies (see Sect.~\ref{sec:spectra}).

If the accretion flow geometry is such that the inner part is occupied by the hot flow 
and the outer is the standard cold disc, the seed photons for Comptonization 
might be internal to the hot flow or come from the outer disc. 
For the truncation radius of the  cold disc of more than 30$R_{\rm S}$ (where $R_{\rm S}=2GM/c^2$ is the Schwarzschild radius),  
most of the disc photons go directly to the observer and therefore  
the disc should be very prominent in the total spectrum. 
Furthermore, the luminosity in disc photons being Comptonized 
in the region of major gravitational energy release ($<10R_{\rm S}$) 
would be only about 1\% of the total luminosity resulting in an amplification factor of a hundred  
and a very hard Comptonization spectrum (see equation~\ref{eq:gamma_delta}). Neither is observed.  
An overlap of the inner hot flow with the cold disc (\citealt{PKR97}; see Fig.~\ref{fig:geom}a)  
was proposed as a solution to this, but does not really solve the problem, because most of the energy is dissipated within 10$R_{\rm S}$. 
Thus the disc should come to radii  below $10R_{\rm S}$. 
Another solution is that cold clouds embedded into the flow 
reprocess hard photons \citep{CFR92,Krolik98,ZPM98,P98} increasing thus the number of seed soft photons. 

Alternatively, the hot flow itself can generate  enough soft photons by 
synchrotron emission of the same hot electrons that emit the X-rays, 
if the electron temperature $T_{\rm e}$ is sufficiently high to overcome the self-absorption problem. 
The spectrum from the hot optically thin advection-dominated accretion flows (ADAF) indeed 
is produced mostly by synchrotron self-Compton (SSC) mechanism \citep{NY94,NY95b,NMQ98,YN14}. 
Detailed radiative transfer calculations accounting for non-local Compton effect  
coupled with dynamics  (also in the  Kerr metric)  predict, however, 
$T_{\rm e}$  exceeding the observed values of 50--100 keV by at least a factor of 2 
(see fig.~3  in \citealt{Yuan07}, fig.~1d in  \citealt{Xie10} and figs.~5 and 6 in \citealt{NXZ12}). 
Most of the problems get solved if the electrons have a reasonably strong non-thermal tail. 
In this case,  synchrotron emission becomes  much more efficient  \citep{WZ01} increasing the cooling, 
softening the spectrum and lowering $T_{\rm e}$  to the values which agree with observations. 
This is the essence of the hybrid Comptonization (or rather  hybrid SSC) models developed for bright accreting BHs 
\citep{PC98,coppi99,PV09,MB09,VVP11,VPV13} that are described below in more details. 
Non-thermal electrons can also play a role in low-luminosity systems such as Sgr A* 
\citep{Mahadevan98,OPN00,YQN03}. However, in these conditions the equilibrium 
electron temperature is very high ($\sim$ MeV) and the optical depth is very low, so that thermal synchrotron radiation is very effective. 
The role of the non-thermal electrons is then reduced to production of tails at higher and lower frequencies 
around the dominating thermal  synchrotron peak.

 \subsection{Hybrid Comptonization model} 
 
\begin{figure}
\centerline{\includegraphics[width=0.46\textwidth]{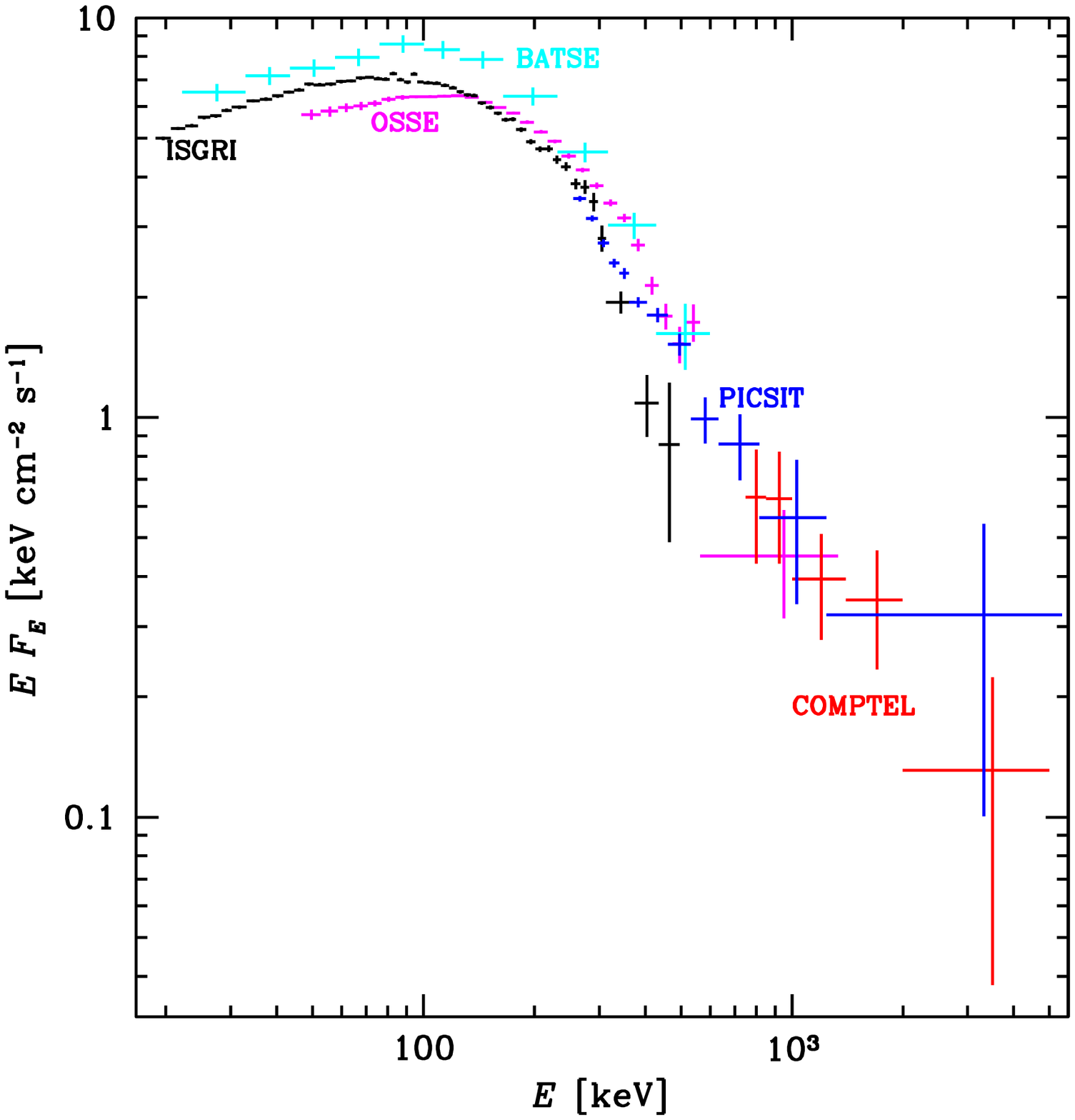}
\hspace{0.2cm}
\includegraphics[width=0.50\textwidth]{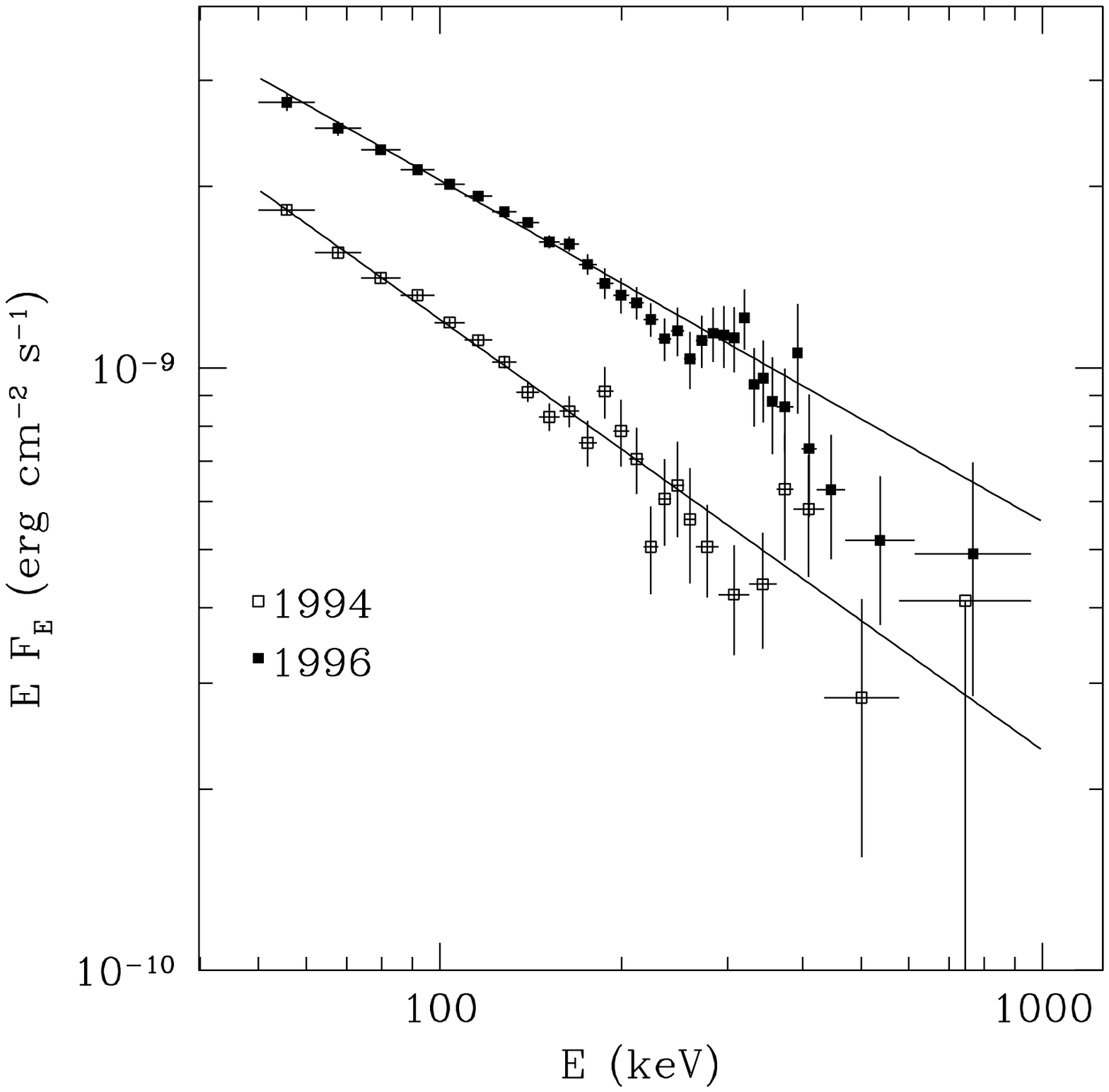}}
\caption{Hard (left panel) and soft (right) state spectra of Cyg X-1. 
From \citet{ZLS12} and \citet{Gier99}. }
\label{fig:tail_cygx1}      
\end{figure}

Arguments in favour of the presence of non-thermal particles  in the accretion flow 
come from the significant detection of the MeV tails in the hard state spectra of Cyg X-1 
\citep[][see Fig.~\ref{fig:tail_cygx1}]{McConnell94,McConnell02,Ling97,JRM12,ZLS12} 
and marginally in GX~339$-$4 \citep{DBM10}.   
Also in the soft state, power-law tails extending to hundreds of keV and up to possibly 10 MeV 
are present \citep[][see Fig.~\ref{fig:tail_cygx1}]{Grove98,Gier99,ZGP01,McConnell02} and are well described by 
non-thermal/hybrid Comptonization  \citep{P98,PC98,Gier99,coppi99}. 
What is the nature of the non-thermal particles is an open question. 

On the theoretical ground, it is expected that the electrons get some of the energy via 
Coulomb collisions with hot, nearly virial protons (as assumed in ADAF models) or diffusive acceleration 
by MHD turbulence, resulting in ``thermal heating", i.e. energy is transferred to the thermal population 
of electrons. 
However, some fraction of the energy can be transferred to them in the form 
of injection of  relativistic electrons by shock acceleration \citep{FB08,DBL09,HBF12},  
magnetic reconnection \citep{DingYuan10,RQSS12,Hoshino13}, or 
electron-position pair production by decay of pions born in proton-proton collisions  \citep{Mahadevan98}. 
Because the microphysics of electron acceleration and heating in the hot flow is not well established 
from first principles, one can use a phenomenological prescription, where some fraction of the total 
power is given to the electrons as heating and the rest of the energy 
is given by non-thermal injection of power-law electrons. Such hybrid thermal/non-thermal models 
are reviewed by \citet{coppi99}. Models with the least number of free parameters are 
either purely thermal or purely non-thermal. Because the first option clearly contradicts the data, 
we consider in the following the second non-thermal option. 
Of course, the assumption that the electrons receive 100\% of their energy in the form of non-thermal injection is not realistic. 
Fortunately, the results are not very sensitive to the actual value of non-thermal injection fraction as long 
as it exceeds 10\% \citep[see Appendix B1 in ][]{VPV13}. 
The reason is that even for pure non-thermal injection the steady-state electron 
distribution is thermal at low energies due to thermalisation via Coulomb collisions between electrons 
as well as via synchrotron self-absorption \citep{GGS88,GHS98,NM98,VP09}. 
At higher energies  a  tail develops, 
whose shape is determined by the injection and 
the competition between various cooling/thermalisation mechanisms.

The most advanced hybrid models solve simultaneously for the momentum distribution of all 
considered particles, electrons and positrons (sometimes also protons), as well as the photons. 
This either can be done via Monte-Carlo simulations \citep[e.g.][]{SBS95}, or by solving coupled 
kinetic equations \citep{Coppi92,coppi99,BMM08,VP09}. 
The processes that need to be accounted for under the conditions of the hot flows  
are Compton scattering, synchrotron emission and absorption, pair production, Coulomb collisions 
(between leptons as well as with protons), and bremsstrahlung.      
The radiative transfer can be easily handled exactly with Monte-Carlo approach, 
while usually  with the kinetic approach an escape probability formalism in a 
single-zone approximation is used. 
A single-zone approximation works reasonably well 
if most of the escaping radiation at all wavelengths is dominated by some narrow range of radii. 
For the X-ray production this is fine, because most of the energy is dissipated in the accretion flow 
spread from say 3$R_{\rm S}$ to 10$R_{\rm S}$. On the other hand, this approximation 
fails in the OIR. Here the outer zones of the hot flow can dominate the energy output in those bands 
as the inner zones are opaque for that radiation. 
Because the radiation from the inner zones can affect the energy balance and the escaping radiation 
from the outer zones, a multi-zone treatment with the radiative transfer is required 
\citep{VPV13}.

The simplest model is  described by the size, the magnetic field strength $B$,  Thomson 
optical depth $\tau$, the total injected power $L$,
the spectrum of the injected electrons and the spectrum and luminosity of the external 
(blackbody/cold accretion disc) photons.  
For an extended multi-zone flow one can assume that the electron energy injection rate  as well as 
$B$ and $\tau$ have power-law distributions with radius, 
$B(R)\propto R^{-\beta}$, $\tau(R)\propto R^{-\theta}$ and thus the 
additional parameters, e.g., $\beta$ and $\theta$ have to be introduced. 
Many parameters  can be directly determined from observations or taken from 
theoretical accretion disc models. 
Now let us describe the main properties of such hybrid models. 

\begin{figure}
\centerline{\includegraphics[width=0.98\textwidth]{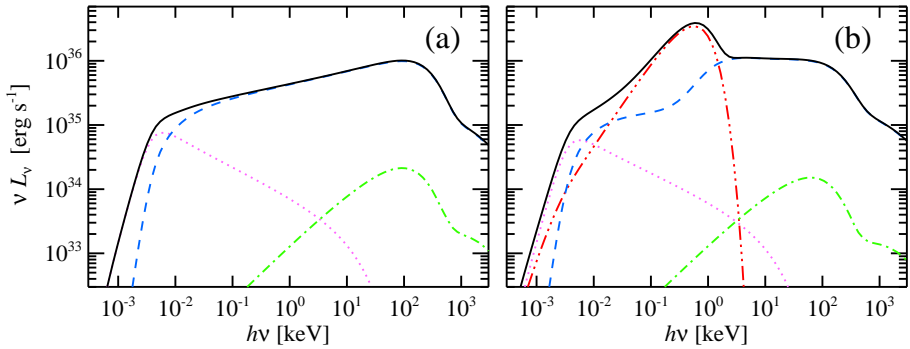}}
\caption{Spectrum from the one-zone hot flow.  
(a) Spectral decomposition of the hybrid SSC model. 
The pink dotted line is the non-thermal synchrotron, the blue dashed is its Comptonization spectrum, 
and the green dot-dashed curve is the bremsstrahlung component. 
(b) Same as (a), but with the disc photons (red triple-dot-dashed curve) dominating 
the seed photons distribution. Now below 1 keV Comptonization spectrum  is produced by SSC, while 
at higher energies Comptonization of the cold disc photons dominates.  }
\label{fig:ssc_decomp}       
\end{figure}

\subsection{Basic properties of hybrid accretion flows} 
 
\label{sec:hyb_hot}

If the truncation radius of the  cold disc is significantly larger than the 
region of the major energy dissipation (i.e. $R_{\rm tr}>30R_{\rm S}$), 
then the X-ray spectrum is dominated by the radiation from the innermost zone of the hot flow.
Here locally generated  non-thermal  synchrotron photons  are Comptonized by the thermal electron population
(see Fig.~\ref{fig:ssc_decomp}a). 
Here we note that the internally generated synchrotron photons are much more 
efficient in cooling the plasma than the external disc photons. The first obvious difference 
comes from the geometry:  all synchrotron photons are injected within the hot flow and have a chance to be Comptonized,
while in the disc case only a small fraction gets to the hot flow. 
The second difference comes from the fact that the synchrotron photons have much smaller energies than 
the cold disc photons. Therefore, in order to produce the same spectral slope of the Comptonization 
continuum (with nearly the same total power),  the synchrotron luminosity can be smaller by a factor a several than the disc 
luminosity intercepted by the hot flow. In other words, the amplification factor given 
by equation~(\ref{eq:gamma_delta}) is now closer to 50 than 10 (because we need to use $\delta\approx 1/10$).   

At a few per cent of the Eddington luminosity corresponding to the bright hard state,  
the low-energy electrons are thermalised by Coulomb collisions  
and synchrotron self-absorption to the typical electron temperatures $T_{\rm e}$ of about 100 keV \citep{PV09,MB09,VVP11}.  
The hybrid model produces surprisingly stable spectra with photon index $\Gamma\sim1.6$--1.8 
largely independent of the model parameters (Fig.~\ref{fig:ssc_decomp}a).  
The high-energy electron tail can be approximated by a power-law, which is softer 
than the injected distribution due  to the cooling. 
The observed  MeV tail is produced by Compton up scattering of the 100 keV photons by these non-thermal electrons. 
The outer zones of the hot flow have softer spectra because of the additional cooling 
by  the cold disc photons and because of more transparent conditions for the synchrotron radiation 
(see Fig.~\ref{fig:ssc_decomp}b). The overall X-ray spectrum is thus concave. 

The OIR spectrum consists of two components: 
the multi-colour (possibly irradiated) cold accretion disc  and the synchrotron radiation from the hot flow. 
Similarly to the inhomogeneous synchrotron models developed for extragalactic jets
\citep{Marscher77,BK79}, 
the non-thermal synchrotron spectrum of the hot flow is a  power-law $F_\nu\propto \nu^{\alpha}$ 
with the index \citep{VPV13}: 
\begin{equation}\label{eq:OIR_slope}
 \alpha_{\rm OIR}    =    \frac {5 \theta + \beta (2p + 3) - 2p - 8}{\beta (p+2) + 2\theta},
\end{equation}
where $p$ and $\theta$ are the indices of the equilibrium distribution of electrons, 
$n_{\rm e}(R,\gamma)\propto R^{-\theta} \gamma^{-p}$,  
at Lorentz factor $\gamma$ emitting at the self-absorption frequency. Typically, spectral indices are $\alpha_{\rm OIR}\sim 0\pm0.5$. 
The turn-over at longer wavelengths is determined by the extent of the hot flow, 
while the transition to the optically thin synchrotron emission is hidden by the Comptonization spectrum (Fig.~\ref{fig:ssc_decomp}a).  
  
\begin{figure}
\centerline{\includegraphics[width=0.47\textwidth]{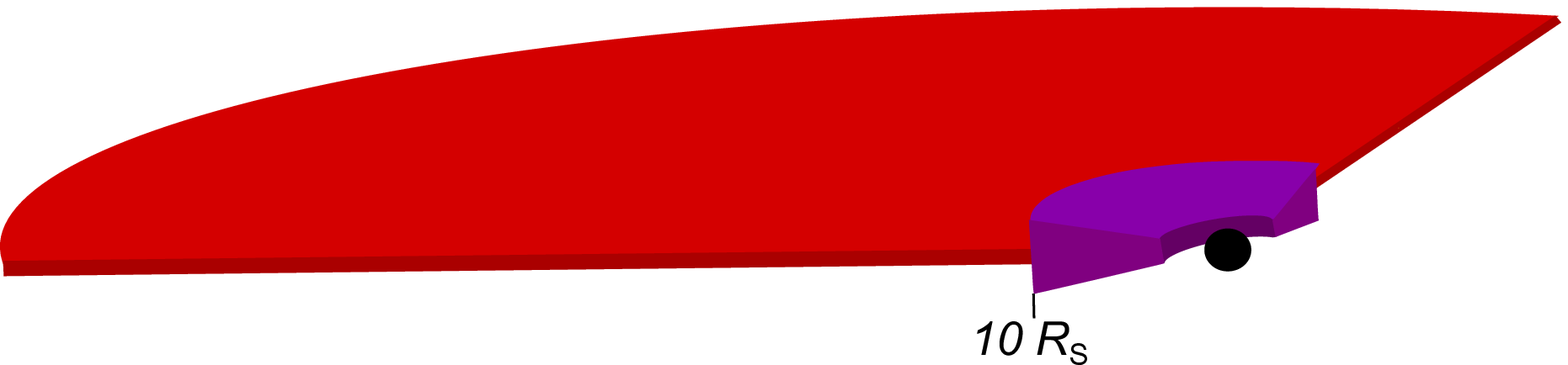}
\hspace{0.2cm}
\includegraphics[width=0.50\textwidth]{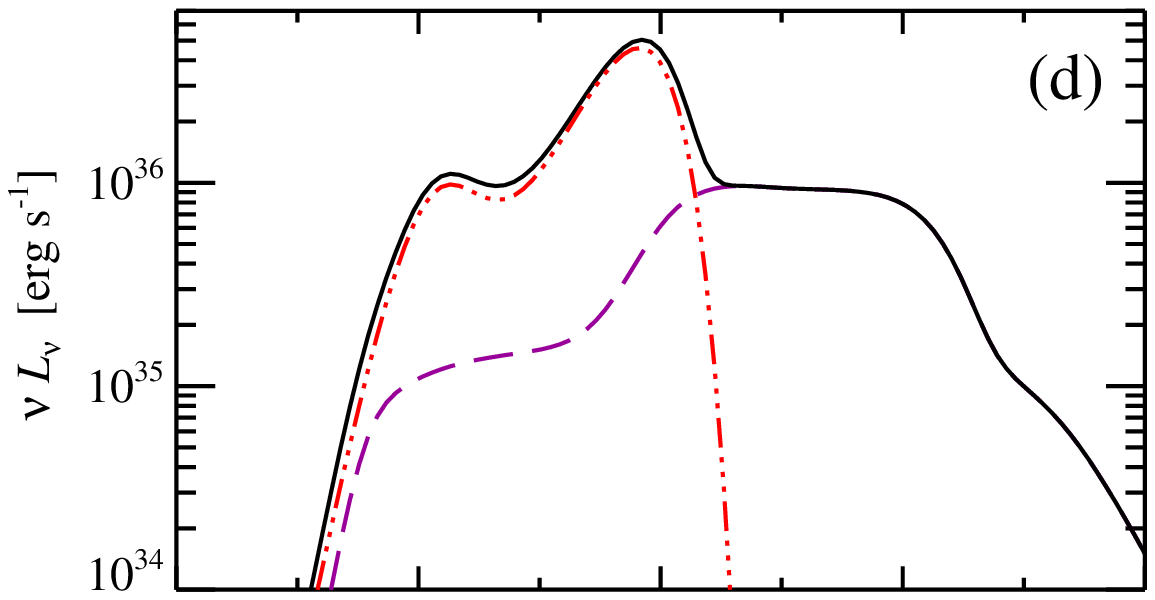}}
\centerline{\includegraphics[width=0.47\textwidth]{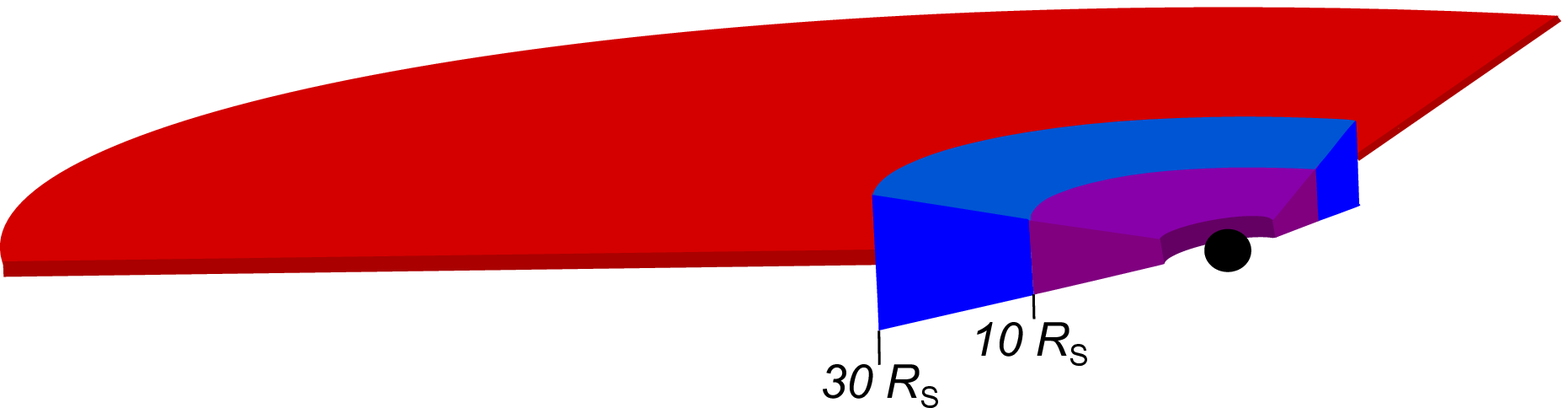}
\hspace{0.2cm}
\includegraphics[width=0.50\textwidth]{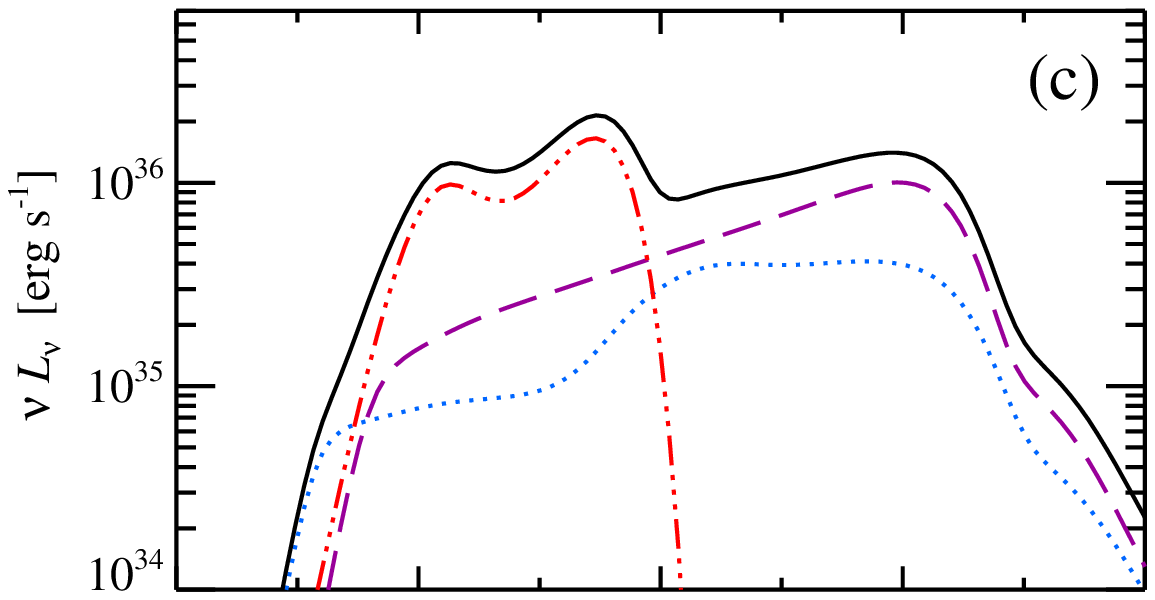}}
\centerline{\includegraphics[width=0.47\textwidth]{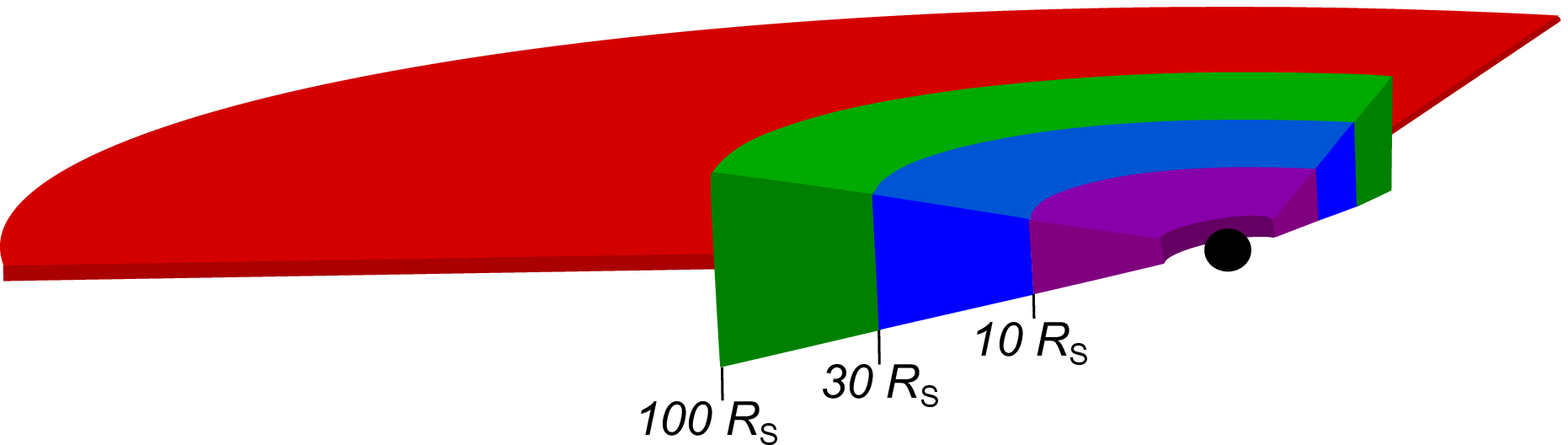}
\hspace{0.2cm}
\includegraphics[width=0.50\textwidth]{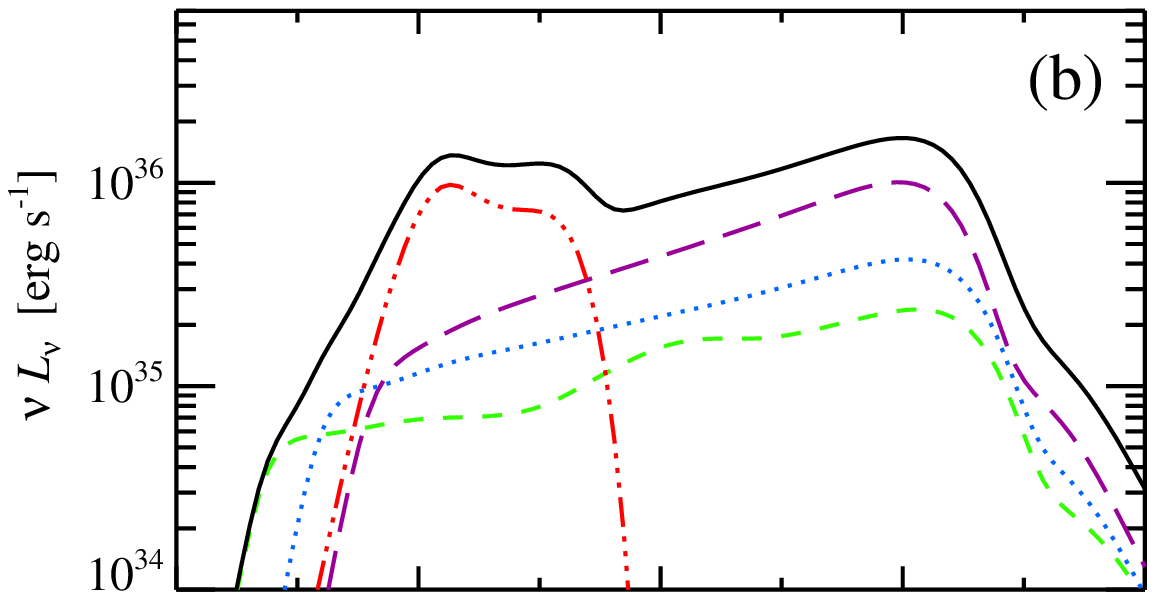}}
\centerline{\includegraphics[width=0.47\textwidth]{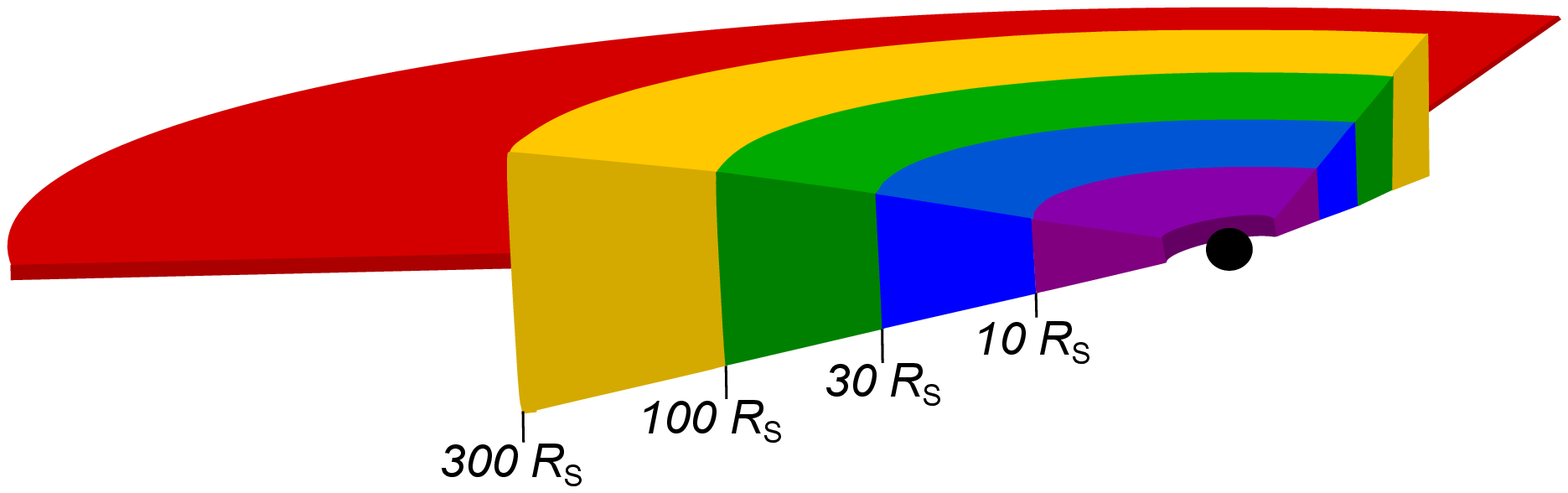}
\hspace{0.2cm}
\includegraphics[width=0.50\textwidth]{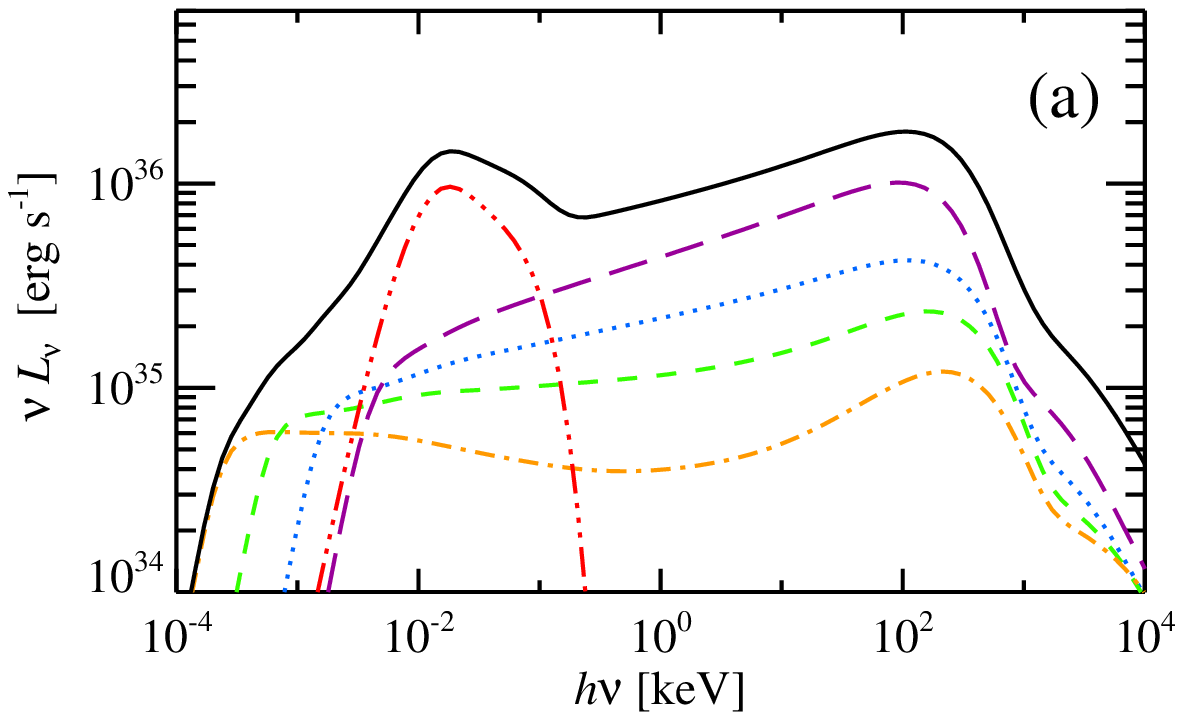}}
\caption{\textit{Left panels}:  
Schematic picture of the evolution of the hot flow  size during spectral state transitions. 
In the low-luminosity hard state, the geometrically thick inner hot flow is large dominating the 
radiative energy output.  With increasing accretion rate (from lower  to  upper panels) 
the outer zones of the hot flow gradually collapse swallowed by the cold thin accretion disc. 
For illustrative purposes, the hot flow is split into four zones with outer radii 
$10R_{\rm S}$ (violet, zone 1), $30R_{\rm S}$ (blue, zone 2), $100R_{\rm S}$ (green, zone 3) and $300R_{\rm S}$ (yellow, zone 4).
Red outer component represents the truncated cold accretion disc. 
\textit{Right panels}: corresponding spectral evolution at the state transition.
Contribution of different zones are marked with different lines: zone~1 (violet long-dashed), zone~2 (blue dotted), zone~3 (green short-dashed), 
zone~4 (orange dot-dashed) and outer cold irradiated disc (red three-dot-dashed).  
At the lower panel, the hot flow spectrum was calculated not accounting for the seed photons from the disc.
An IR excess is clearly visible above the irradiated disc spectrum. 
Collapse of the hot flow leads to dramatic changes in the OIR hot flow synchrotron spectrum. 
Changes in the X-ray spectral shape are insignificant until the truncation radius becomes as small as $10R_{\rm S}$. 
The Comptonization spectrum from the hot-flow zone closest to the cold  disc 
consists of two separate continua produced by Comptonization of the synchrotron and the cold disc photons, 
with the latter being dominant source of seed photons in this zone  (see also Fig.~\ref{fig:ssc_decomp}b). 
For simplicity, in this illustration the total luminosity is kept constant. 
Adapted from \citet{VPV13}.  
}
\label{fig:trans_spectra}       
\end{figure}

The disc spectrum can be split also into two components: the inner warmer standard disc 
heated by viscous forces and the outer cooler disc heated by the X-rays. 
In the OIR one expects the dominance of the irradiated disc, which 
has the radial temperature dependence  $T_{\rm irr} \propto R^{-3/7}$ \citep{Cun76}.  
Presence of the irradiated disc can be  reflected in the optical echoes 
\citep{Hynes98,OBrHH02}, 
the X-ray time-lags \citep[see][and references therein]{Pou02} and 
in the optical/X-ray cross-correlation function \citep{HBM09,VPV11}. 
Its signatures are also seen in the spectrum \citep[e.g.,][]{HHC02,GDP09}.
For typical parameters of LMXBs with the disc size of $10^{11}$~cm and the X-ray luminosity of $10^{37}$~erg~s$^{-1}$ 
the temperature of the outer disc is about $T_{\rm irr} \sim 20\ 000$~K. 
The relative role of the components varies with the wavelength.  
The disc spectrum is hard in the OIR band, while the hot flow produces an excess emission 
dominating below $\sim$1~eV (see Fig.~\ref{fig:trans_spectra}a and  Fig.~\ref{fig:broad-band-gx339-4}).

At smaller accretion rates below a few percent of the Eddington value, 
the flow becomes more transparent to the synchrotron 
photons leading to their increasing role in cooling and resulting in slightly {\it softer} X-ray spectra  \citep{VVP11}.  
At higher accretion rates associated with the transition to the soft state (see Fig.~\ref{fig:trans_spectra}), 
the outer zones of the hot flow gradually collapse, so that the 
disc truncation  radius $R_{\rm tr}$ decreases \citep{PKR97,Esin97}.
This leads to the rising role of the disc as a source of seed photons, which
increases Compton cooling, leads to spectral softening 
and causes changes in the electron distribution from mostly thermal to nearly non-thermal \citep{PC98,PV09,MB09,VVP11}. 
This transition is accompanied by the increase in the  reflection amplitude
that scales with the solid angle at which the cold disc is seen from the hot flow.
The OIR hot flow luminosity drops first at longer wavelengths, where the outer zones radiate (see Fig.~\ref{fig:trans_spectra}). 
Note that the X-ray spectrum changes much later, when the truncation radius comes closer 
to the zone of the main energy dissipation of about 10$R_{\rm S}$. 
At this moment, the Comptonization spectrum of the remaining hot flow consists of two segments: 
the hybrid SSC dominates at lower energies, while Comptonization of the disc photons takes over  
at energies above the cold disc peak (see Fig.~\ref{fig:ssc_decomp}b). 
Similarly curved spectra are expected for the hot flow zones closest to the cold disc (Fig.~\ref{fig:trans_spectra}). 
The corresponding time delay between sharp luminosity changes at different wavelengths 
scale with the timescale of state transition and,  
depending on the separation of the wavelengths and accretion parameters, 
can be as short as hours (e.g., if one observes in different optical filters),  
as long as a few days (e.g., IR and UV) or weeks (e.g. IR and X-rays). 
The opposite evolution should be observed in the soft-to-hard spectral transition 
when the accretion rate drops after the outburst peak. 
Here first the X-ray spectral transition starts and at a timescale of a week or so 
the emission  in the OIR peaks, when the size of the hot flow becomes large enough for the OIR synchrotron photons to escape.

At a high accretion rate,  the accretion disc extends to the last stable orbit and the source switches to the soft state. 
In this case, no  inner hot flow exists, but a non-thermal magnetically powered corona  still  could be present (Fig.~\ref{fig:geom}b). 
Its presence is supported by the existence of the X-ray/$\gamma$-ray power-law tails. 
The non-thermal synchrotron from the corona may be present in the OIR band, but at a much lower level, 
because  the electron cooling is dominated now by Comptonization of the disc photons.
Actually, such a non-thermal corona atop the cold disc (in addition to the hot inner flow)
may be present also in the hard state, but its emission scaled with the cold disc luminosity is weak. 
 
 \begin{figure}
\centerline{\includegraphics[width=1.0\textwidth]{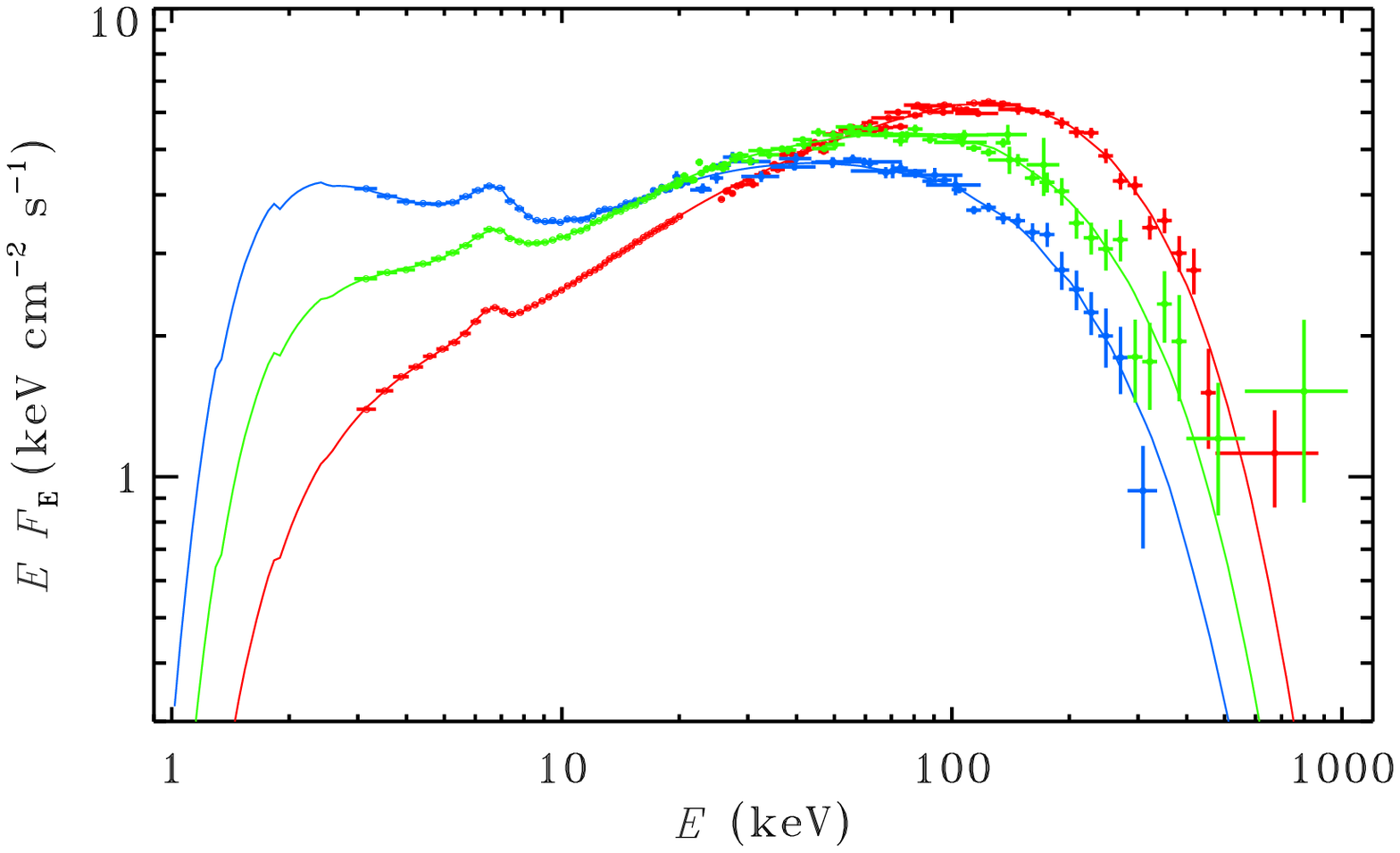}}
\vspace{0.5cm}
\centerline{\includegraphics[width=1.0\textwidth]{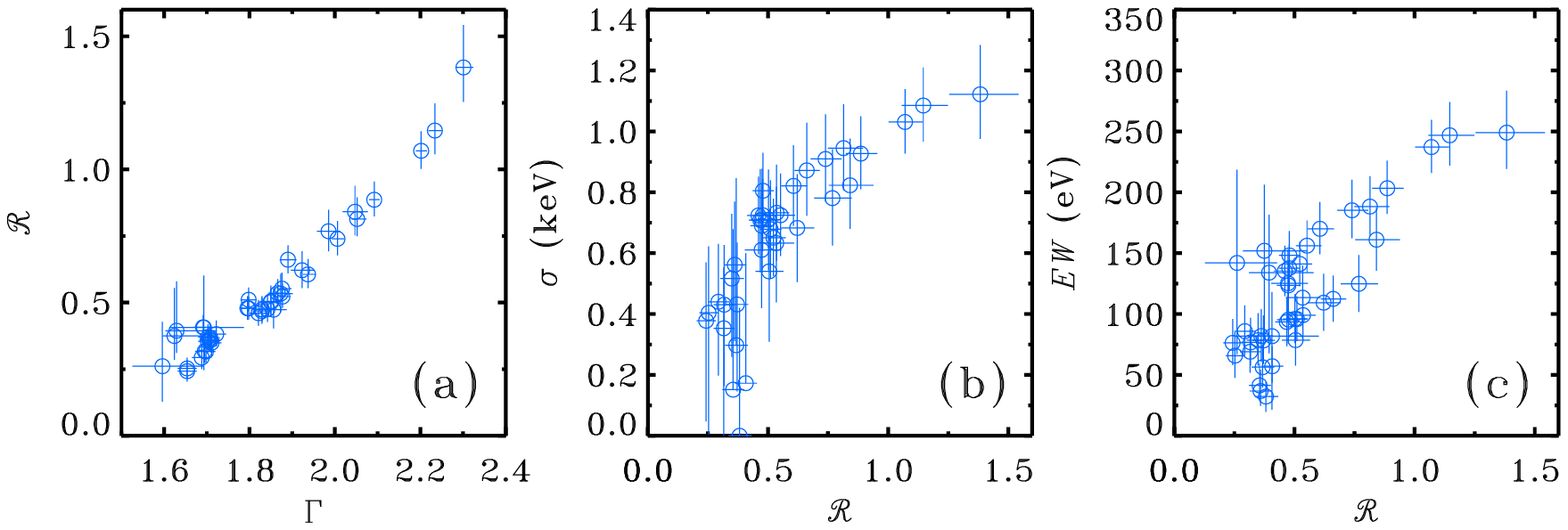}}
\caption{Spectral variations of Cyg X-1 in its hard/intermediate state as shown in the top panel.  
The lower panels show correlation between the amplitude of Compton reflection $\cal{R}$, 
photon  spectral index $\Gamma$, width  $\sigma$ and the equivalent width of the iron line.
Adapted from \citet{IPG05}. }
\label{fig:hard_cygx1}      
\end{figure}

\section{Observational properties} 
\label{sec:data}

\subsection{X-ray/$\gamma$-ray spectra}
\label{sec:spectra}
 
Let us first briefly summarise what we know about the spectral properties of BHBs. 
Further details can be found in reviews by \citet{P98}, \citet{ZG04}, \citet{DGK07}, and \citet{Done10}. 
In the hard-state, the spectra constitute a power-law in the X-ray band 
with a rather stable spectral slope (with photon index $\Gamma\sim 1.6-1.9$ 
and ubiquitous sharp cut-off at  around 100~keV (\citealt{G97,ZPM98,IPG05}; see fig.~10a in \citealt{ZG04}). 
The shape of the spectra  allows us to conclude that they 
are produced by (nearly) thermal Comptonization \citep[e.g.][see Fig.~\ref{fig:hard_cygx1}]{P98,ZG04}. 
When the high-quality data above 100 keV were available 
(e.g. with OSSE/\textit{CGRO} or IBIS/\textit{INTEGRAL} or HXD-GSO/\textit{Suzaku}) 
the electron temperature  (measured with the accuracy of about 10\%)\footnote{Note that the 
electron temperature can be measured only if high-quality data are available above 100 keV and 
accurate Comptonization models such as {\sc compps} \citep{PS96} or { \sc eqpair} \citep{coppi99} are used for 
fitting. Using the exponentially cut-off power-law for the fits and 
identifying the e-folding energy with the electron temperature is dangerous, 
because that model does not correctly describe  the shape of Comptonization continuum 
\citep[see e.g. fig.~5 in][]{ZLG03}. This can result in over-estimation of $T_{\rm e}$ by a factor of 3--6. }
was always  lying in the interval 50--120 keV 
\citep[e.g.][]{G97,ZPM98,P98,ARLS04,MTYD08}, with temperature increasing with decreasing luminosity. 
The hard-state accreting BHBs also show weak but distinctive MeV tails 
\citep[][see Fig.~\ref{fig:tail_cygx1}]{McConnell94,McConnell02,Ling97,DBM10,JRM12,ZLS12}.
Such spectra are consistent with the hybrid hot-flow  model. 
During the transition to the soft state, $T_{\rm e}$ is reduced. However, 
because of the growing importance of non-thermal tail, the spectral cutoff energy may actually increase \citep{PV09}.

The fact that the spectra are stable in the hard state with variable luminosity and 
never have $\Gamma<1.6$ argues in favour of the hybrid SSC as the main emission mechanism. 
If the outer cold disc were the seed photons  provider, one expects harder and strongly variable 
spectra when the truncation radius is changing in the soft-to-hard transition.  
Moreover, the best studied BHs, Cyg~X-1 and GX~339$-$4, clearly have a concave spectrum 
that can be fitted with two Comptonization continua \citep{FPZ01,IPG05,MTYD08,SUT11,YMD13}. 
The inhomogeneous hot flow model naturally explains such spectra by
the radial dependence of the slope of Comptonization spectrum. 
The spectral curvature can also appear if 
a non-thermal corona (similar to that producing power-law tail
in the soft state) exists above the cold disc during the hard state too \citep{IPG05}.  
 
In addition to the smooth continuum, in both states  
a Compton reflection feature and the fluorescent iron line at 6.4 keV 
originating from cool opaque matter (likely the cool accretion disc) are often detected.
The strength of Compton reflection is correlated with the X-ray slope \citep{ZLS99,ZLG03}, 
with the width of the iron line as well as with the quasi-periodic oscillation (QPO) frequency 
\citep[][see lower panels in Fig.~\ref{fig:hard_cygx1}]{GCR99,RGC01,IPG05,Gilfanov10}. 
During the outbursts of BH transients, in the hard state  the iron line width 
correlates well   with the luminosity \citep{KDDT13}. 
These data are consistent with the hot-flow paradigm where all 
correlation are basically controlled by the cold disc truncation  radius.  
 
At luminosities above a few per cent of Eddington, BHBs show a strong correlation between spectral index and luminosity.
At lower luminosities the trend is reversed, i.e. spectra become softer with decreasing luminosity \citep{WuGu08,SPDM11}. 
Similarly, an indication of the reverse trend was detected in low-luminosity AGNs \citep{Const09,Gu09}. 
This was interpreted as a change of the source of seed photons for Comptonization from the disc photons 
dominating at higher luminosities to the synchrotron at lower luminosities.  
The whole spectral   index -- luminosity dependence is well explained by   the hybrid 
hot flow model \citep[see figs~7 and 12 in][]{VVP11}.  
At very small luminosities, the flow becomes more transparent for the synchrotron radiation 
increasing the photon input and softening the Comptonization spectra.

\begin{figure}
\centerline{\includegraphics[width=0.88\textwidth]{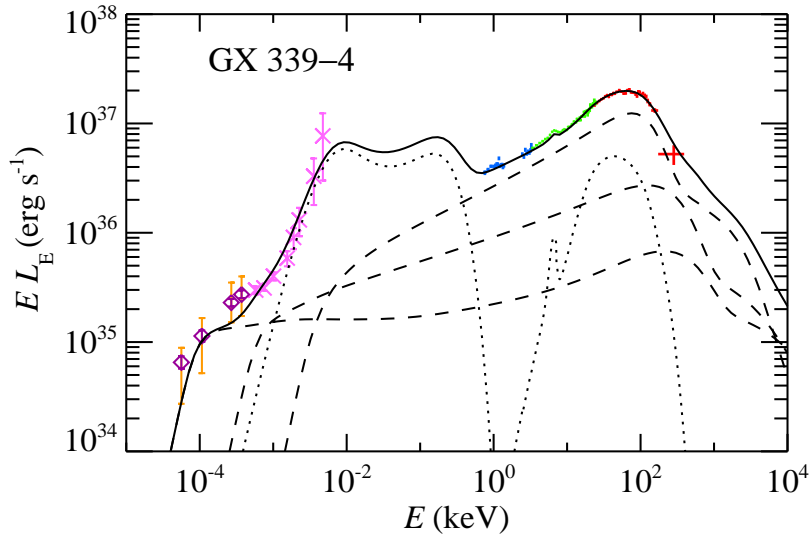}}
\caption{Broad-band spectrum (corrected for absorption) of GX~339$-$4 in its hard state 
around March 5, 2010 from the mid-IR to the hard X-rays  \citep{CB11}. 
The dashed lines show contribution of different zones of the  hot accretion flow \citep{VPV13}, 
the dotted lines represent the  spectra of the irradiated and standard  discs as well as Compton reflection. 
From Veledina et al., in prep. }
\label{fig:broad-band-gx339-4}      
\end{figure}

\subsection{Broad-band spectra and infrared flares} 

Numerous multiwavelength campaigns were conducted over the past decade. 
Broadband radio to X-ray spectral energy distributions (SEDs) for many BHBs were constructed
\citep[e.g.][]{HMH00,MHG01,CHM03,Cadolle07,CB11,DGS09}.  
The OIR emission is normally dominated by the (irradiated) disc, but the IR excesses are observed 
in a number of sources: XTE~J1859+226 \citep{HHC02}, 
GX~339$-$4 \citep{GBRC11,SUT11,Buxton12,DKB12,RCC12}, 
A0620$-$00 \citep{GMM07}, SWIFT~J1753.5$-$0127 \citep{CDSG10}, V404 Cyg \citep{HBR09}, 
XTE~J1550$-$564 \citep{Jain01b,RMDF11}. 
In some cases the OIR spectrum can be described by a pure power-law  $F_\nu\propto \nu^{\alpha}$ with index $\alpha$ close to zero 
(e.g. $\alpha_{\rm OIR}=-0.15$ in  XTE~J1118+480, see \citealt{Esin01,CHM03}). 
Sometimes the OIR excess spectrum is rather soft with $\alpha\approx -0.7$ \citep[see][for a recent overview]{Kalemci13}. 
The IR excesses were previously explained by the jet \citep{HHC02,GMM07} 
or the dust heated by the secondary star \citep{MM06}.
\citet{VPV13} recently  argued  that the OIR excess emission may also be  produced by  
synchrotron radiation from the hot flow.

\begin{figure}
\centerline{\includegraphics[width=0.80\textwidth]{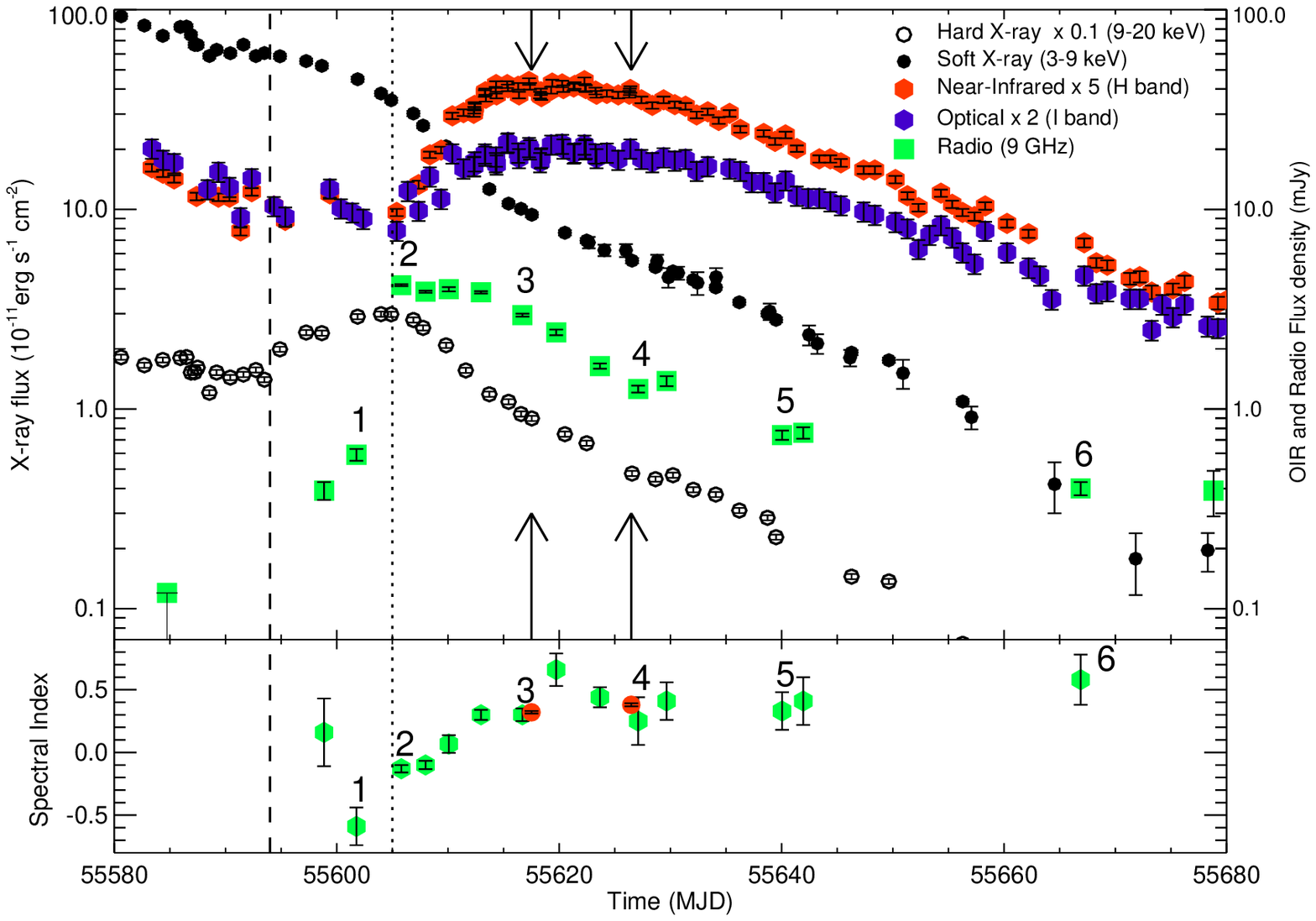}} 
\centerline{\includegraphics[width=0.80\textwidth]{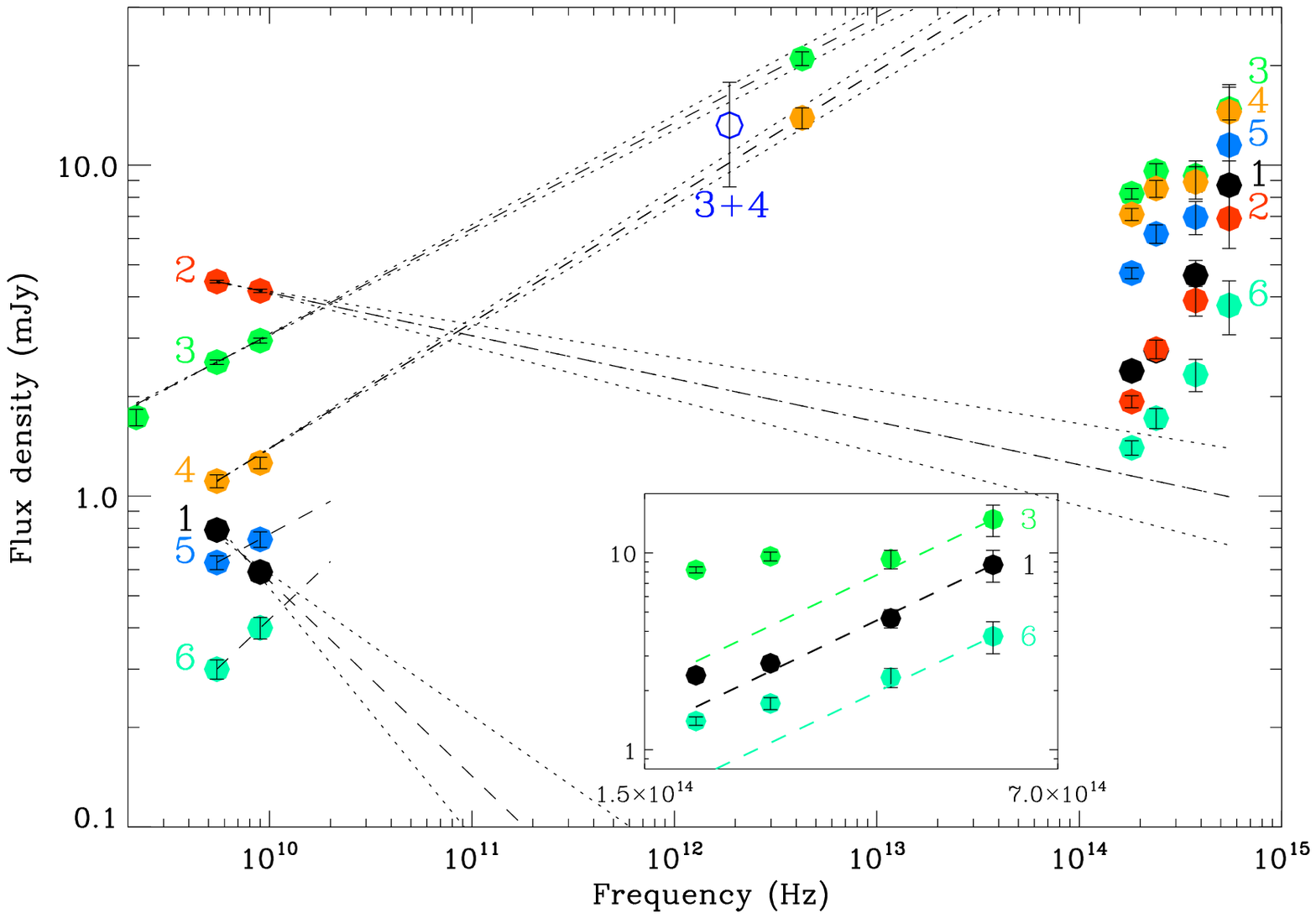}} 
\caption{Upper panel: The lightcurves  of GX~339$-$4
during the decay of the 2010--2011 outburst in radio (green), 
NIR (red), optical (blue), soft X-ray (filled black circles) and hard X-ray (open black circles). 
The bottom sub-panel shows the evolution of the radio spectral index. 
The dashed vertical line  indicates the start of the  soft-to-hard state transition, while the 
dotted line  marks the start of the OIR flare and the end of the transition \citep[see fig.~1 in][]{DKB12}. 
The numbers refer to the individual observations displayed in the lower plot.
Lower panel: Evolution of the radio to OIR spectra of GX~339$-$4. 
The inset zooms on the OIR spectra, with the dashed lines corresponding 
to power-law with $\alpha=1.5$ (for an irradiated  disc).
From \citet{Corbel13}.  }
\label{fig:gx339-4_corbel}
\end{figure}

Here we only discuss a few representative examples of the recent studies. 
The broad-band spectrum of GX~339$-$4 in its hard state \citep{CB11,CD13IAUS} is shown in  Fig.~\ref{fig:broad-band-gx339-4}. 
The X-ray spectrum peaking  at $\sim$100 keV  is well described  by thermal Comptonization. 
The irradiated disc presumably dominates in the UV band.
There is a clear excess in the mid- and near-IR \citep{GBRC11}, but the spectrum becomes harder at longer wavelengths. 
These data are well explained by the non-thermal synchrotron emission from the hot flow \citep{VPV13}
of about 500$R_{\rm S}$.

Excellent data covering both radio and OIR bands have been collected during 
the 2010--2011 outburst of GX~339$-$4 \citep[][see Fig.~\ref{fig:gx339-4_corbel}]{CB11,DKB12,Corbel13}. 
A week after  the start of the transition to the hard state (marked by the vertical dashed line)
the OIR spectrum has a clear soft excess above the reprocessing thermal emission (point 1), 
while the radio jet optically thin emission can contribute at most 1\% to the OIR.    
Few days later, when the transition was completed (point 2), the fluxes 
in the $H$ and $I$-band  show a sharp increase and an obvious IR excess in the OIR spectra 
is visible, while the radio  emission is still soft with  $\alpha\sim -0.1$.
On a week time scale the radio  spectrum transits to the  harder, optically thick state 
with $\alpha\sim 0.5$ (point 3) corresponding to the  synchrotron emission 
from an inhomogeneous source analogously to the extragalactic jets \citep{BK79}.
The radio spectrum stays hard and the IR excess is visible during the following decay. 
The  high flux in the {\it Herschel} far-IR band lies exactly on the extrapolation of the radio spectra. 
Because after the break, the jet spectrum must be optically thin and soft, while the 
OIR spectra are flat (even after  subtracting thermal component, 
as discussed by \citealt{DKB12,Buxton12,Corbel13}), the jet does not contribute significantly  to the OIR bands. 
An important conclusion from these data is that 
there is a rather strong, evolving component in the OIR band which cannot be produced by the jet or reprocessing
in the accretion disc at any stage of the outburst. 
Its appearance in the hard state is, however, consistent with the hot flow interpretation.

\begin{figure} 
\centerline{\includegraphics[width=\textwidth]{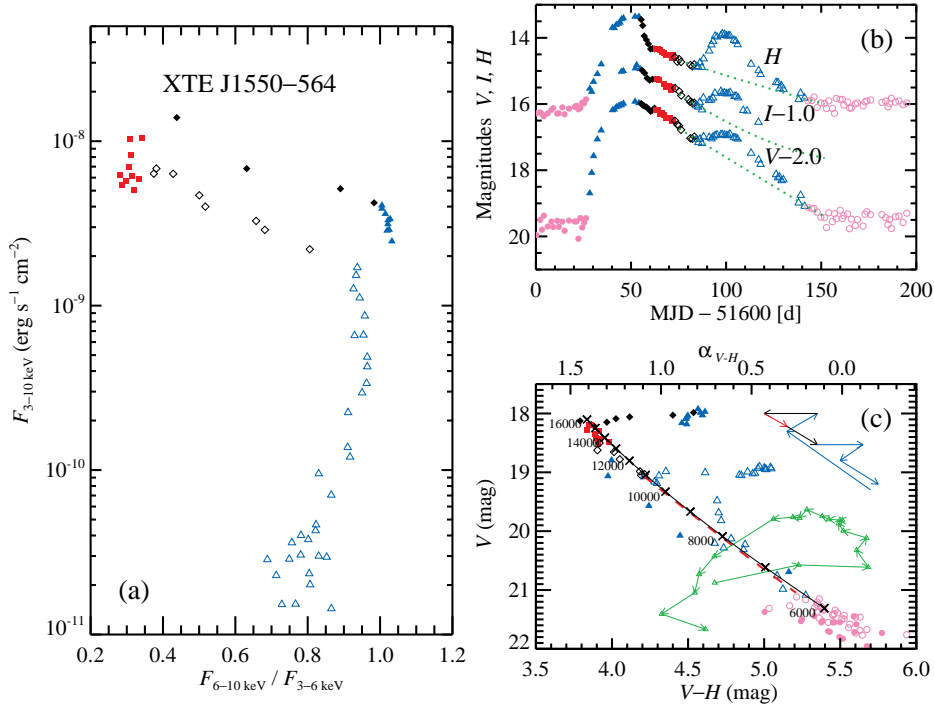}}  
\caption{(a) X-rays flux -- hardness diagram  for XTE~J1550$-$564 
during the 2000 outburst. Different colours indicate various stages of the outburst 
based on their X-ray hardness. 
The blue colour indicate the hard state, the black colour is the transition and the red colour 
the soft state. Open and closed symbols correspond to the rising and decaying outburst stages, respectively. 
(b) The light curves in $V, I$ and $H$-filters  of XTE~J1550$-$564 \citep{Jain01b}. 
The pink circles correspond to the quiescent state. 
The green dotted curves show the best-fit model for the decaying disc component.  
(c) The observed $V$ vs $V-H$   colour-magnitude diagram.
The black solid line represent the relation expected for the black body disc of a characteristic radius $2.7\times10^{11}$ cm 
inclined at $i=75^{\rm o}$ at distance of 4.38 kpc \citep{Orosz11} of different temperatures (marked next to the line). 
The model magnitudes were reddened following the extinction law of \citet{Fitzpatrick99} with  $A_V=5.0$. 
Much smaller $A_V$ (that would lead to softer spectrum) 
is not possible, because the disc temperature would be below hydrogen ionisation temperature 
needed for the outburst to start. 
The blue-black-red arrows illustrate schematically the time evolution of the source 
during the outburst. 
The green arrows show the path followed by the flare after MJD 51680 (see the text). 
The upper x-axes show the intrinsic spectral index $\alpha_{VH}= 4.63 - 0.84\ (V-H)$ 
computed from the observed colour. 
We see that the flare component is never softer than $\alpha=-0.2$. 
From Poutanen et al., in preparation.
}
\label{fig:xte1550}
\end{figure}

The IR excess  similar to that seen in GX~339$-$4 in the hard state appears also in 
XTE~J1550$-$564 \citep{Jain01b}, 4U~1543$-$47 \citep{BB04} and XTE~J1752$-$223 \citep{Russell12}. 
The properties of the excess can be studied using the colour-magnitude diagram (see Fig.~\ref{fig:xte1550}c).
During the 2000 outburst of XTE~J1550$-$564 the data in the soft state, soft-to-hard transition 
and at very low-luminosity hard state can be adequately described 
by the reddened irradiated disc emission.  
Both at the rising (filled symbols) and decaying  (open symbols) 
phase of the outburst, we see ``flares" during which the spectrum becomes redder. 
One can interpret these flares as the appearance of the additional red component.
Fitting the fluxes at the decaying stage with an exponential plus a constant,
one can subtract the contribution from the irradiated disc and obtain a spectrum of the flare component only. 
We see that the second flare starts with the spectral index $\alpha\sim+0.7$, then becomes 
softer with $\alpha\sim-0.2$, and then harder again (see green arrows in Fig.~\ref{fig:xte1550}). 
(The final points have large errors, because of the uncertainties in the subtraction of the disc.)
What is important that the flare starts in $I$ before $H$, 
so that the index measured between filters $I$ and $H$ is even larger, $\alpha_{IH}\approx1.0$.
This behaviour rules out immediately the interpretation of the flare  
in terms of optically thin jet emission.\footnote{Colour-magnitude diagram 
for the 2000 outburst of XTE~J1550$-$564 was constructed by \citet{RMDF11}, who 
also related the observed colours to the intrinsic spectral indices 
and claimed very soft spectrum of the flare. 
Unfortunately, all their formulae are wrong for various reasons and 
the actual intrinsic spectra are much harder.  
Furthermore, the exponential fits to the OIR light curves to evaluate the flare spectrum  
were also flawed, as their fits overestimate the disc contribution 
in the $V$ and $I$ filters just before the flare (see fig. 2 in \citealt{RMDM10})  
resulting in over-subtraction of the flux in those filters and 
in a much too soft spectrum of the flare 
(compare our $\alpha\sim+0.7$ at the start of the flare with their $\alpha\sim-1.6$).  } 
Instead the data are consistent with the  inhomogeneous hot accretion flow model of \citet{VPV13}. 
Such indices were also observed in the flare spectrum of XTE~J1752$-$223 \citep{Russell12}
and GX~339$-$4 \citep{DKB12,Buxton12}. 
This would mean that all these OIR flares are produced by the hot flow but not the jet. 
The observed sharp colour change during the flares is related to the collapse/recover of a 
zone in the hot flow that is responsible for the H-band emission. 

The delay of the IR flare peak by about 10 days from the start of the soft-to-hard transition  
is naturally expected, because the X-ray transition corresponds to the start of the 
retraction of the cold disc ($R_{\rm tr}\sim 10 R_{\rm S}$), 
while the IR flare peaks when the hot flow is large enough ($R_{\rm tr}\gtrsim 100 R_{\rm S}$) allowing 
the IR photons to escape (see Fig.~\ref{fig:trans_spectra}). 
At the rising phase of the outburst, just a few day before the hard-to-soft spectral transition, 
a dip has been observed in the UV light curve of GX 339$-$4  \citep{YY12}. 
In  Swift~J1910.2$-$0546 a dip first appears in the IR, then optical and finally in the UV  (N. Degenaar, priv. comm.). 
The timing of the dips is consistent with the collapse of the hot flow with increasing accretion rate (see Sect.~\ref{sec:hyb_hot}). 

It is worth noticing that the hard-to-soft and the soft-to-hard spectral transitions occur at different X-ray luminosities 
\citep[e.g.][]{ZGM04}. This hysteresis is most probably related to the fact that  at the same luminosity 
the cold disc is further away from the central source on the rising phase of the outburst, than on the decline. 
The hysteresis should be then also reflected in the OIR spectra, namely 
the fast colour change  should occur at a higher X-ray luminosity on the  rising phase, than on the decline, 
as indeed observed.

Some BHBs, however, do show signatures of the jet emission in the OIR band. 
The most obvious examples is microquasar GRS~1915+105, 
whose radio light-curve was found to be very similar to the IR one with a few hours delay \citep{FPB97} 
favouring a common origin. It, however, accretes at a nearly Eddington rate and  is hardly representative. 
The OIR spectrum of 4U~1543$-$47 and MAXI~J1836$-$194 
in the hard state is rather soft with $\alpha\sim -0.7$ \citep{Kalemci05,Russell13}, 
which is consistent with the optically thin synchrotron emission from the jet. 
Furthermore, the rms spectrum of the IR variability of XTE J1118+480 during the 2005 outburst is  close to a
power-law with $\alpha\sim -0.8$ \citep{HRP06}, implying probably the jet origin. 
And finally, GX~339--4 in the hard state demonstrated strong correlated IR and X-ray variability with 
the IR lagging by 0.1~s, which could be interpreted as a signature of propagation delays between the X-ray producing  
accretion disc and the jet \citep{CMO10}. 
It is well possible that three components (the irradiated disc, 
the hot flow and the jet) contribute to the OIR band and their contribution can vary not 
only from source to source, but also in the same source from the outburst to the outburst.

\subsection{X-ray variability} 
\label{sec:xray_variab}

In addition to the spectral properties, the variability in X-rays and longer 
wavelengths puts strong constraints on the models. 
In  the hard state, the typical power-density spectra (PDS) that describe the X-ray variability 
can roughly be  represented as a doubly-broken power-law 
with indices  0, $-1$ and  $-2$ from low to high frequencies.   
A more accurate description of the PDS is achieved by representing it 
with the Lorentzians  \citep[e.g.][]{Nowak00,ABL05}.  
The main source of the short-term variability in BHs is believed to be fluctuations in the mass accretion rate, 
propagating through the accretion flow \citep{Lyub97}.

\begin{figure}
\centerline{\includegraphics[width=0.50\textwidth]{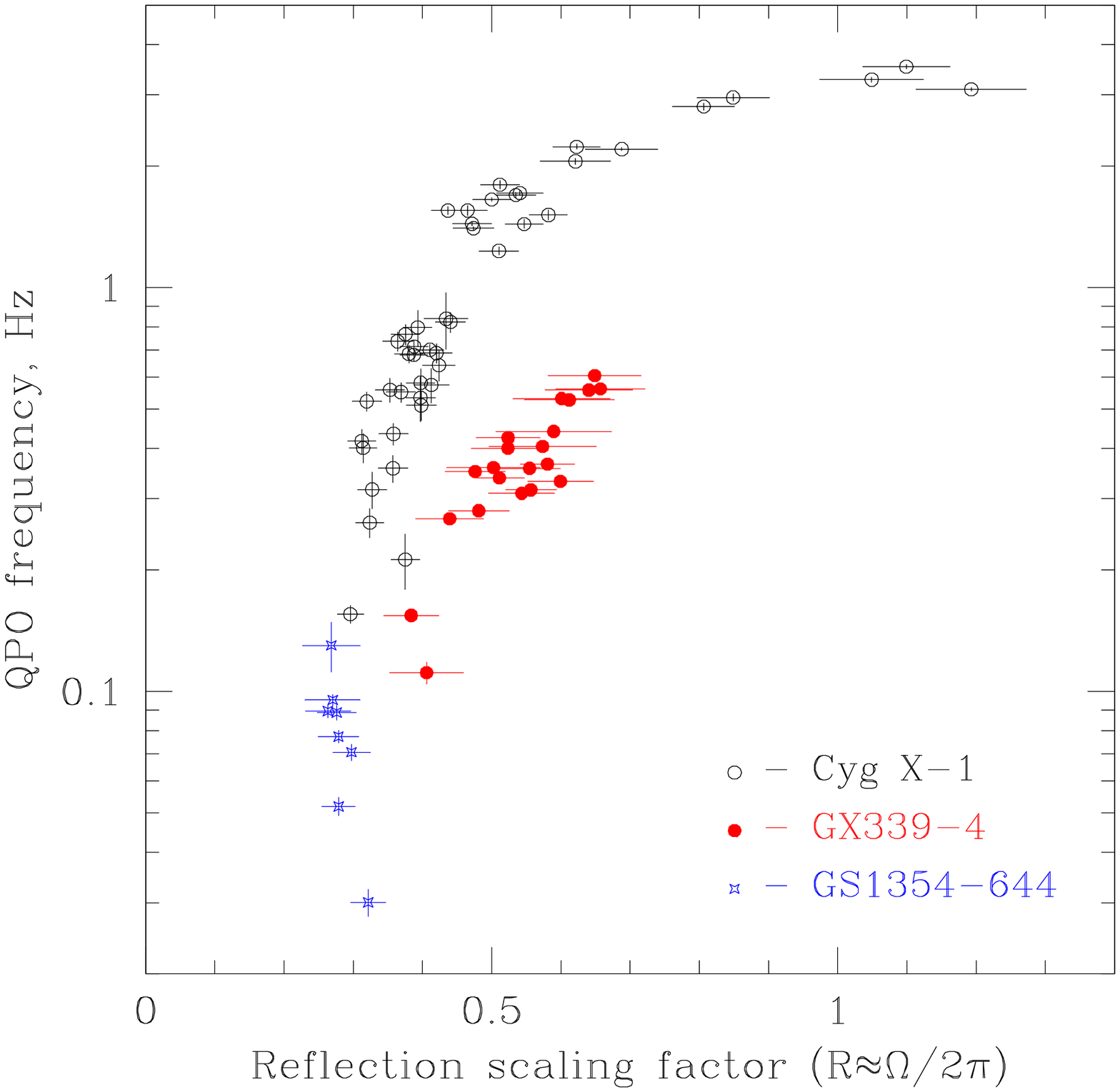}
\includegraphics[width=0.50\textwidth]{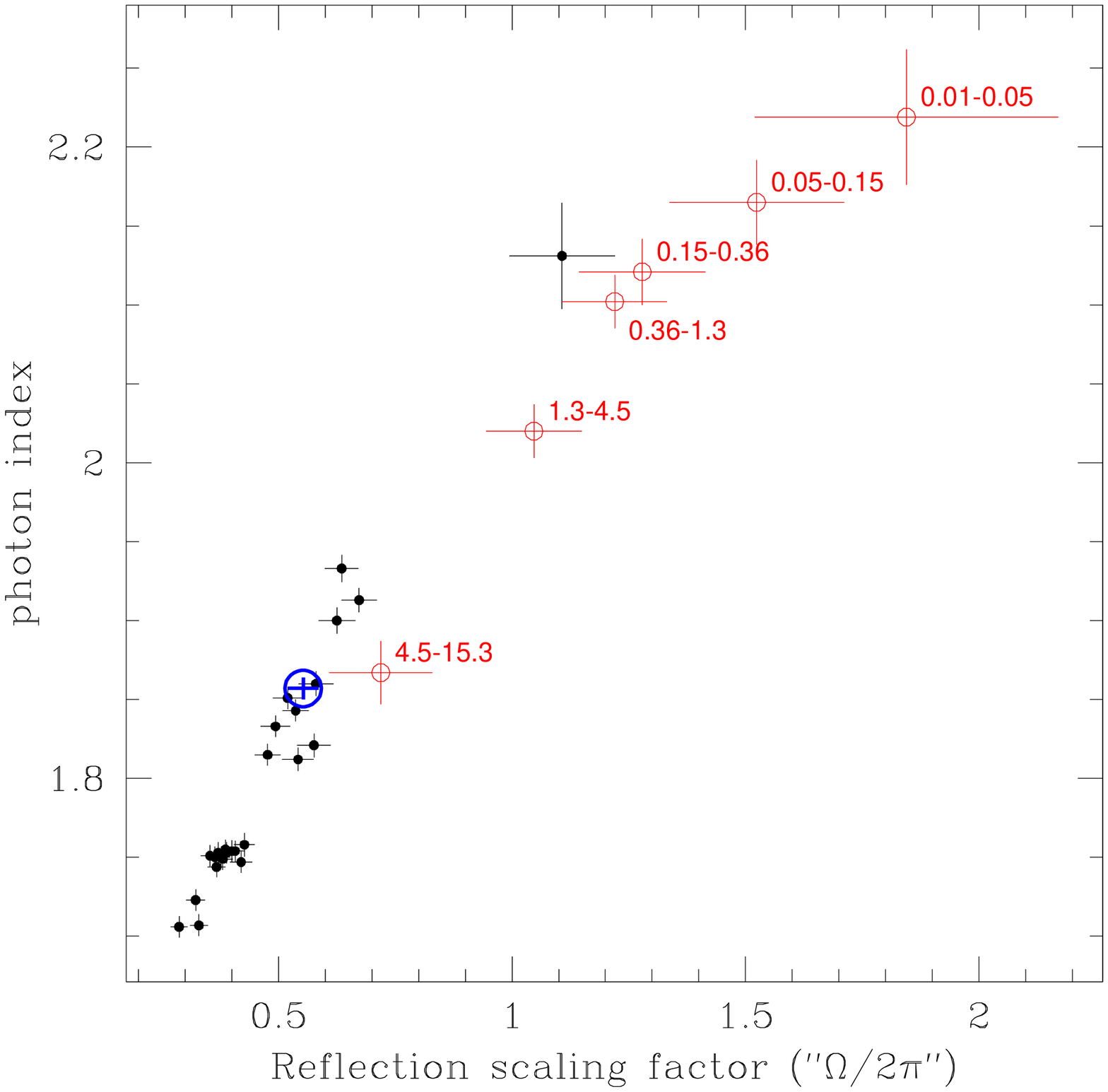} 
}
\caption{Left: QPO frequency vs reflection correlation. From \citet{Gilfanov10}. 
Right:   Photon index --   reflection correlation in the average spectra 
(same as Fig.~\ref{fig:hard_cygx1}a)
and in the Fourier frequency resolved spectra (FFRS) at different frequencies. 
From \citet{GCR99}. 
}
\label{fig:rgamma_cygx1}      
\end{figure}

Often QPOs are observed in the range $\sim$0.1--10~Hz  \citep[see reviews by][]{RM06,DGK07}. 
Their frequencies show correlation  with the X-ray flux, 
amplitude of Compton reflection and anti-correlation with the hardness ratio (see Figs~\ref{fig:hard_cygx1} and 
\ref{fig:rgamma_cygx1}).  
The origin of QPOs is often associated with the  precession of orbits around the BH due to a 
misalignment of the BH and the orbital spins, known as  Lense-Thirring precession  
\citep[e.g.,][]{SV98}, or with oscillation modes of the accretion flow itself \citep[e.g.,][]{WSO01}.
The problem with the Lense-Thirring precession models is that the frequency for test masses is a 
strong function of radius and the BH spin \citep[see e.g.][]{SHM06}. 
It is not clear why any specific radius gets selected to produce a QPO. 
If that radius is defined by the truncation of the cold disc, why the QPOs are 
then observed in the Comptonization spectrum? 
However, if the flow is hot and thick, it will precess as a solid body \citep{FB07} 
with the frequency mostly depending on its size (which is a function of the accretion rate) and weakly on the BH spin and 
the flow height-to-radius ratio. 
In such a case, the precession frequencies lie in the observed range 
and  the model explains well their correlations with other quantities \citep{IDF09,ID11}.

\begin{figure}
\centerline{  \includegraphics[width=0.78\textwidth]{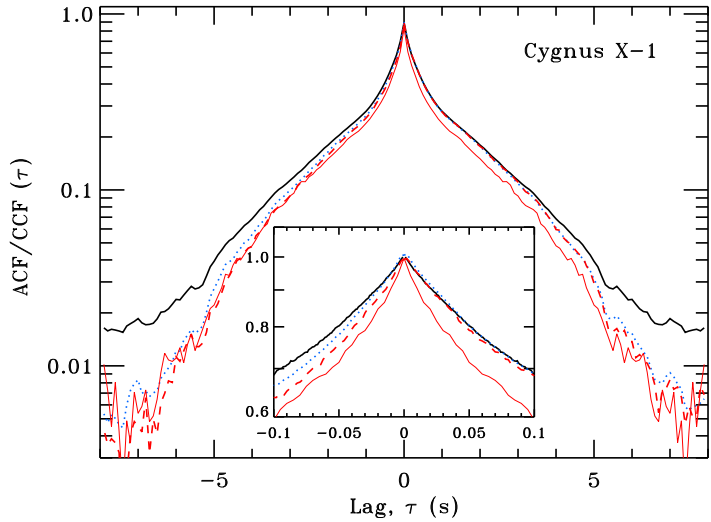}}
\caption{The ACF and CCFs of  Cygnus  X-1  observed  in December  1997.  
Solid  curves  show the ACFs for the 2-5 keV  energy  band
(black solid  curves) and the 24-40 keV band (red solid  curves).  
The blue dotted  curve shows the CCF  between the 8-13 keV and the 2-5 keV bands, 
and the red dashed  curve represents  the CCF for the 24-40 keV  vs the 2-5 keV  bands.  
The positive lag corresponds here to the hard photons lagging the soft ones.  
The CCFs are asymmetric, but the peak do not show any shift from zero lag. 
From \citet{MCP00}.}
\label{fig:acf_cygx1}     
\end{figure}

\begin{figure}
\centerline{\includegraphics[width=0.55\textwidth]{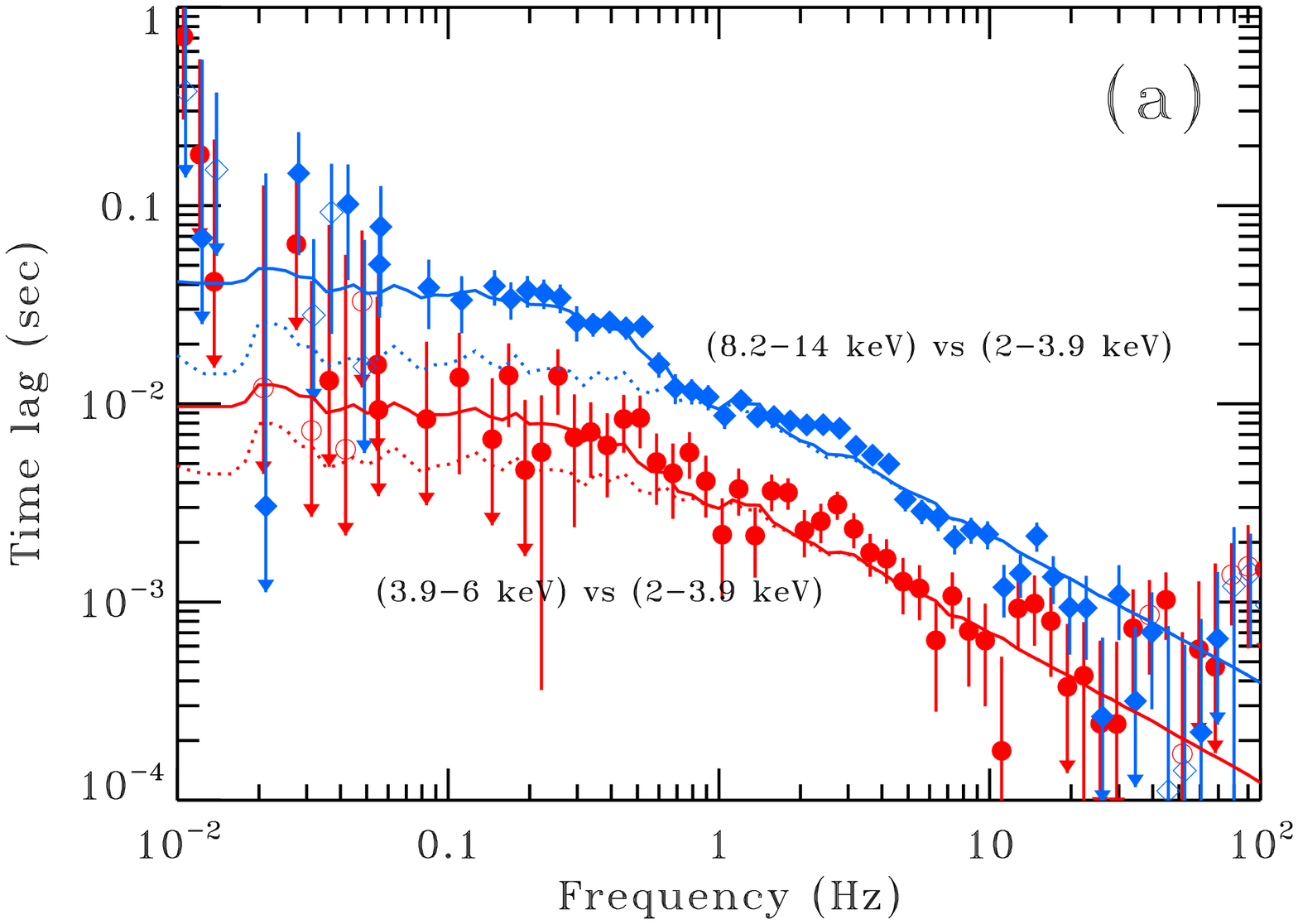} 
\includegraphics[width=0.45\textwidth]{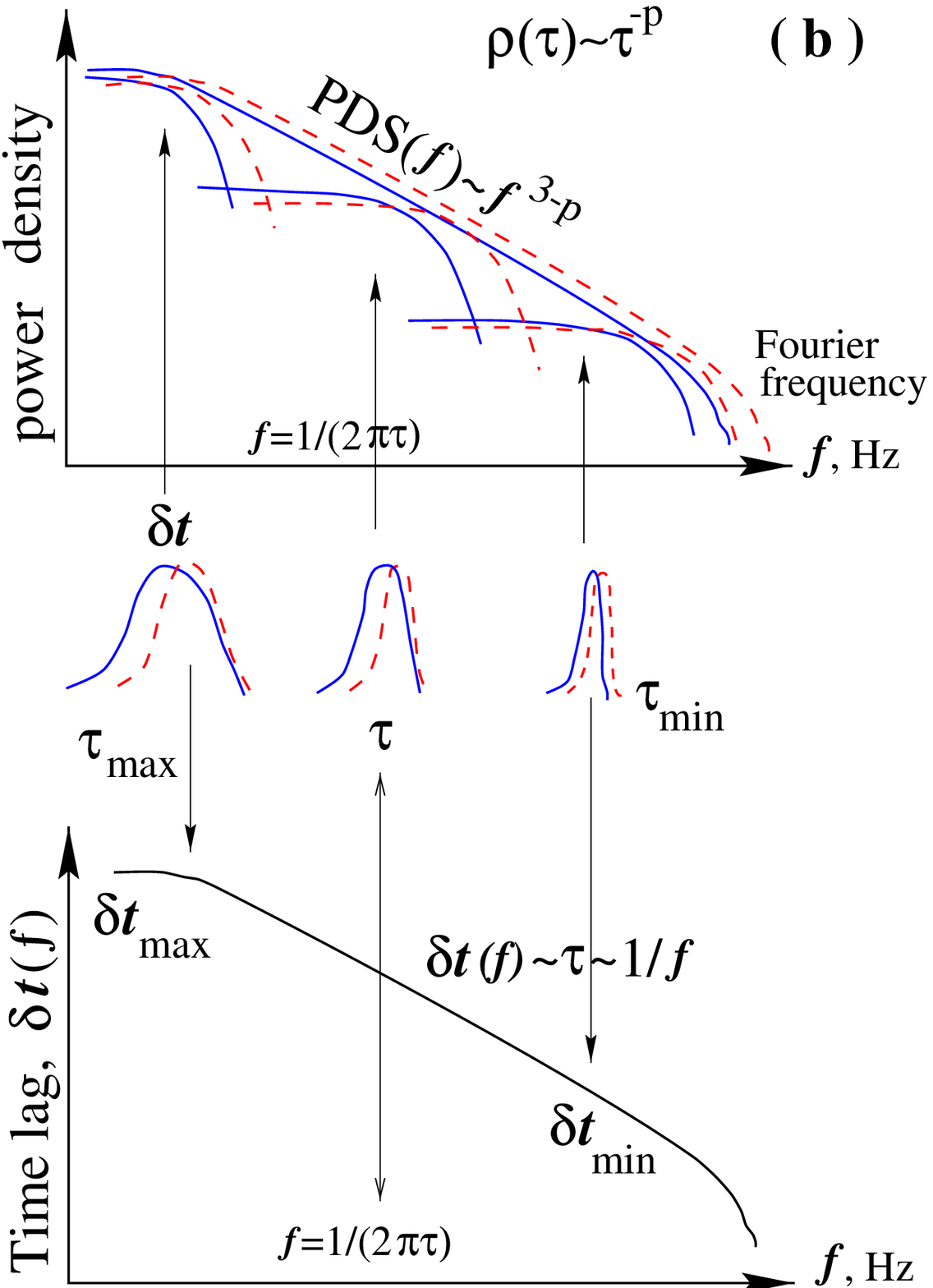}
}
\caption{(a):  Time lags in Cyg X-1. The data points are from \citet{Nowak99a}. 
The  dotted curves represent the model of \citet{PF99}, where lags are produces 
by spectral evolution during flares. The solid curves 
have a contribution  at low frequencies from the lags caused by reflection from the outer disc \citep{Pou02}. 
(b): A schematic picture how spectral evolution model produces time lags. 
The red dotted curves represent the flare light curve at high photon energies, while 
the blue solid curves represent  those at low energies. Spectral hardening during flares 
produces hard time lags. 
Here $p$ is the index of the power-law probability distribution  of flare duration $\tau$.
From \citet{Pou01}. 
}
\label{fig:lags_cygx1}      
\end{figure}

Another way of looking at the variability properties is through the 
autocorrelation function (ACF, which is related to the PDS via Fourier transform). 
The ACF becomes narrower at higher X-ray energies \citep[][see Fig.~\ref{fig:acf_cygx1}]{MCP00}, 
which is equivalent to the excess variability at higher frequencies in the PDS at those energies. 
Moreover, the light curves at different energies are well correlated with each other, 
but the harder X-rays are delayed with respect to the soft X-rays \citep[][see Fig.~\ref{fig:lags_cygx1}]{MiKi89,Nowak99a,Nowak99b}.  
This effect is also reflected in the  asymmetries of the cross-correlation function (CCF) between the 
hard and the soft X-ray energy bands \citep[][see Fig.~\ref{fig:acf_cygx1}]{PGR79,NGM81,MCP00}.

In order to have a better understanding for these asymmetries, it is useful to look at the 
time lags $\Delta t$ between the light-curves at these X-ray energies as a function of the Fourier frequency $f$.   
For the hard state of Cygnus X-1 they are shown in Fig.~\ref{fig:lags_cygx1}a. 
As a function of energy $E$, the lags relative to energy $E_0$ follow the logarithmic law $\propto \ln(E/E_0)$.  
The rather large lags (exceeding 0.1 s) were first interpreted as produced by Comptonization 
in a large Compton cloud \citep{KHT97}. 
For harder photons more scatterings  is required, thus they spend more time in the 
medium before escape  and therefore are delayed. 
Such an interpretation not only causes a problem  with the energetics of the cloud 
(requiring large energy release at distances $>10^4 R_{\rm S}$), 
but also contradicts the energy dependence of the ACF width \citep{MCP00}. 
On the other hand, the large lags, their frequency-dependence $f^{-1}$ 
and the logarithmic energy dependence 
can  be naturally explained by spectral pivoting of a power-law-like spectrum 
if the characteristic time-scale of the evolution scales with the duration of shots $\tau$
dominating variability at frequency $f \approx 1/(2\pi \tau)$ 
 \citep[see][and Fig.~\ref{fig:lags_cygx1}b]{PF99,Pou01,KCG01,KF04}. 
Among the first physical models explaining the spectral evolution 
was the flaring magnetic corona model of \citet{PF99}.  
The observed linear relation between the flux and the rms \citep{UMH01}, however, 
argues against the independent shots (flares) as the source of variability. 
In the propagation model of \citet{Lyub97}, 
spectral evolution can arise when the accretion rate fluctuations 
propagate towards the BH into the zone with the harder spectra  
\citep{MiKi89,KCG01,AU06}.\footnote{For the
time lag production, this model is, however, mathematically identical to the flare evolution model.}
This is consistent with the multi-zone hot flow model,  
because a lower flux of soft seed photon from the cold accretion disc as well as a stronger synchrotron self-absorption 
in the inner part of the flow produce harder  Comptonization spectrum \citep{VPV13}.

Additional contribution to the time lags are possible when the intrinsic X-rays  from the hot flow 
are reflected from the distant matter, e.g. outer cold accretion disc \citep{KCG01,Pou02}. 
Due to the fluorescence and the energy dependence of the photo-electric opacity 
the contribution of  reflection to the total spectrum is energy dependent. 
Thus the Fe line at 6.4 keV and the Compton reflection bump above 10 keV should show 
excess lags. Surprisingly, the deficit of lags (or ``anti-lags'') relative to the logarithmic dependence 
has been observed at these energies  \citep{KCG01}.  
Clearly, this deficit cannot be caused by a simple reverberation and the light travel time effect. 
Instead it could be explained within  the hot flow paradigm as follows. 
The outer parts of the flow which are close to the cold disc 
have softer spectra and larger reflection amplitude,
while the inner flow produces harder spectrum with low or no reflection. 
Thus in the propagating fluctuation model, the reflection will be \textit{leading} the 
hard spectrum causing negative delays at photon energies where it contributes  \citep{KCG01}. 
Delays due to reprocessing in the inner part of the cold disc are seen at high $f$ 
in GX~339$-$4 \citep{UWC11},  while at low $f$ the disc photons lead the Comptonized photons, 
which is consistent with the propagation of fluctuations from the disc to the hot flow.

\begin{figure}
\centerline{\includegraphics[width=0.48\textwidth]{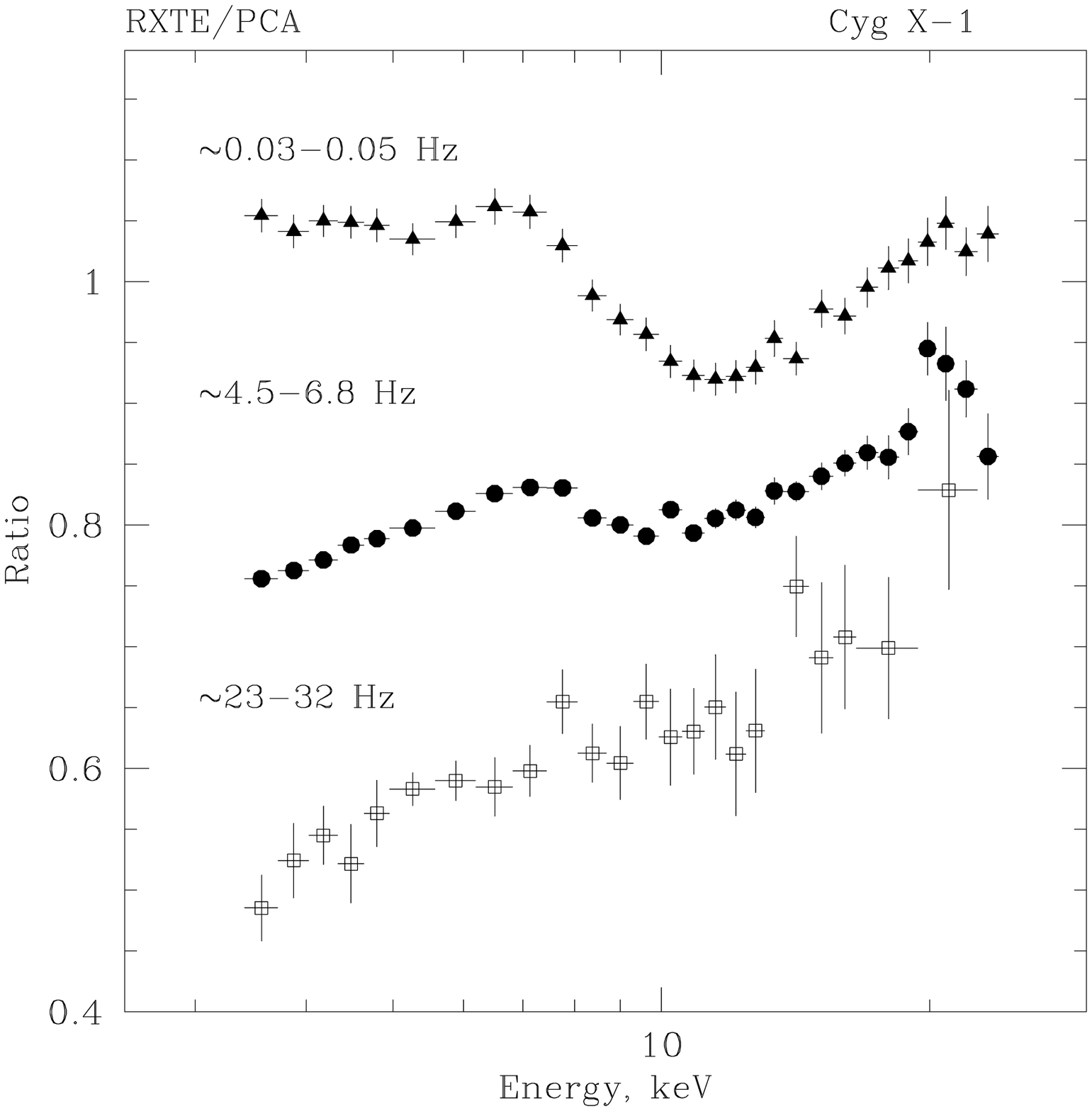} 
\hspace{0.5cm}
\includegraphics[width=0.455\textwidth]{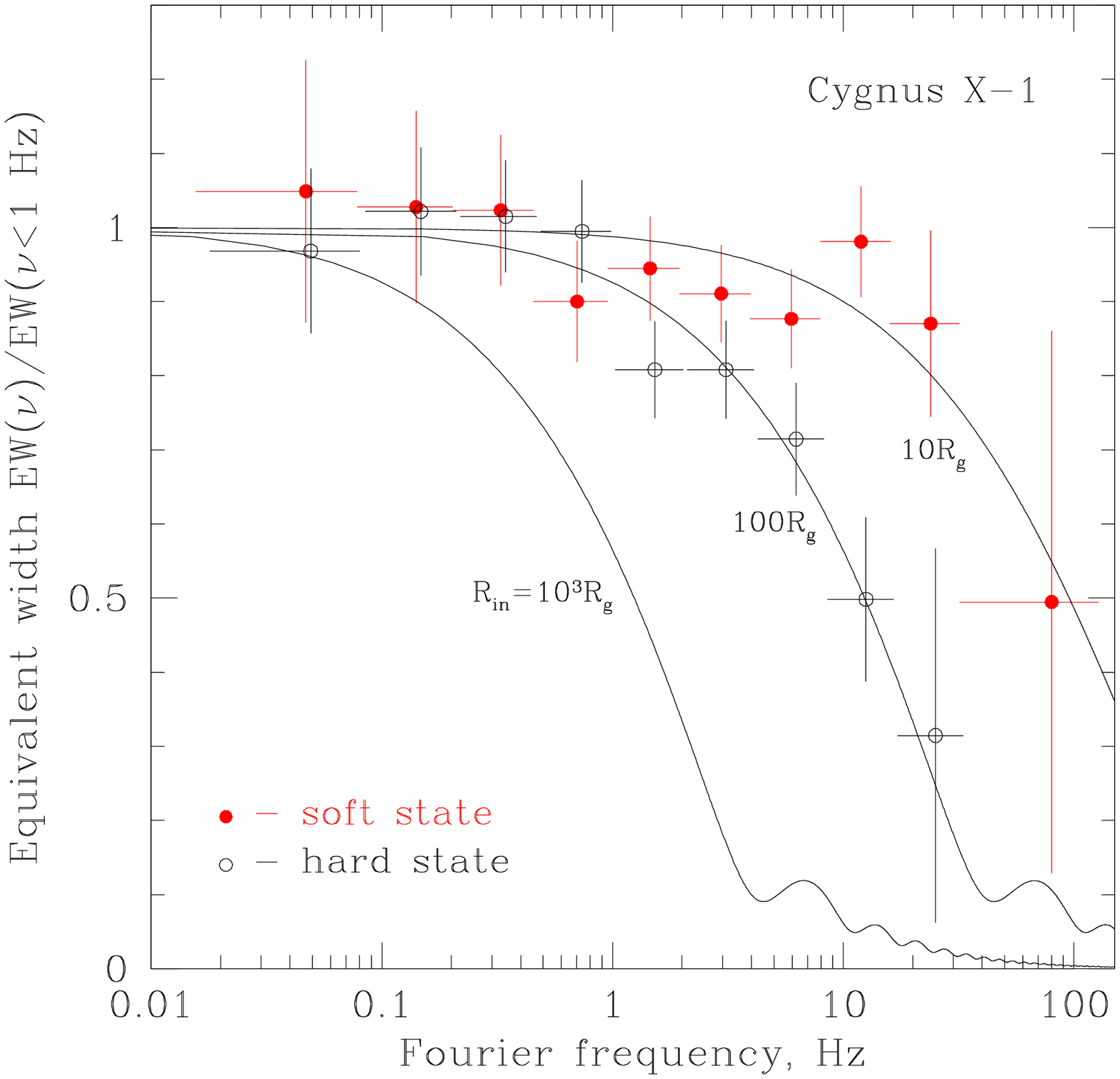}}
\caption{Left: FFRS of Cyg X-1. From \citet{RGC99}.  
Right: equivalent width of the iron line 
as a function of Fourier frequency as measured from the FFRS. From \citet{GCR00}. 
}
\label{fig:ffrs_cygx1}      
\end{figure}

Further support to the truncated cold disc -- hot inner flow scenario comes from 
the Fourier-frequency-resolved spectra \citep[FFRS; ][]{RGC99,GCR00}, 
which are softer and have larger reflection amplitude at low Fourier frequencies (see Fig.~\ref{fig:rgamma_cygx1}, right  
and Fig.~\ref{fig:ffrs_cygx1}, left). 
This implies that soft X-rays are mostly produced  in the outer zones of the hot flow, closer to the cold reflecting medium,
while the hard X-rays are produced in the inner zones that vary at high Fourier frequencies.  
The reduction of the equivalent width of the 6.4~keV Fe line in the FFRS above 1~Hz 
(Fig.~\ref{fig:ffrs_cygx1}, right) 
suggests that the cold disc truncation radius in the hard state is about 100~$R_{\rm S}$ \citep{RGC99,GCR00}. 
Even larger truncation radius of the cold disc (300--700~$R_{\rm S}$) was 
measured in the low-extinction BH transient XTE~J1118+480 \citep{CHM03,Yuan05}, 
where the peak is clearly seen in the UV. 
In the soft state instead the truncation radius is small ($<10~R_{\rm S}$).

\subsection{Optical (IR and UV) variability and its relations to the X-rays}

\begin{figure}
\centerline{\includegraphics[width=\textwidth]{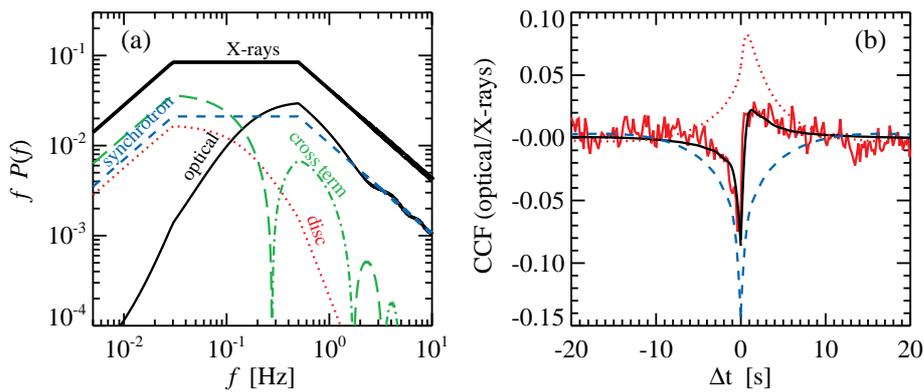}}
\caption{Two-component model for the optical/X-ray CCF. 
(a) The PDSs of the X-ray (upper double-broken line) and optical light curves (thin solid line). 
Three terms contributing to the optical PDS are also shown: synchrotron (blue dashed), reprocessing in the disc (red dotted), and the cross term 
(green dot-dashed  -- positive contribution;  long-dashed -- negative contribution). 
(b) The optical/X-ray CCFs observed from Swift~J1753.5$-$0127 in 2008 
\citep[red noisy curve, see fig. A3 in][]{DSG11}. 
The model CCF is shown by the black solid curve, 
while contributions of the synchrotron and the reprocessed emission 
is  shown by blue dashed and red dotted lines, respectively. 
Adapted from \citet{VPV11}.}
\label{fig:ccf-opt-xray}      
\end{figure}

Data on the fast variability are now available not only in the X-rays, but also at lower energies.
The first simultaneous observations in the optical and X-rays were carried out already 30 years ago by \citet{Motch83} for GX~339$-$4.
Although no confident conclusion could be reached because of the short duration of observations, 
the optical/X-ray CCF  revealed a complicated structure with a precognition dip (i.e. anti-correlation) at negative lags 
corresponding to optical leading the X-rays and a peak (i.e. correlation) at positive lags (see Fig.~\ref{fig:ccf-opt-xray}b). 
Recently, similar CCFs were obtained from the much longer duration simultaneous observations in three BHBs: 
XTE~J1118+480 \citep{Kanbach01,HHC03,MBSK03}, Swift~J1753.5$-$0127 \citep{DGS08,DGS09,DSG11,HBM09} 
and {GX~339$-$4} \citep{GMD08,GDD10}. 
These data provide an important information on the interrelation between various components and give clues to their physical origin.

The observed CCF shape  cannot be explained by a simple reprocessing model  \citep{Kanbach01,HHC03}.  
However, if the optical emission consists of two components, e.g., one coming from the non-thermal 
synchrotron in the hot flow and another from reprocessed X-ray emission, 
the complex shape can be reproduced  \citep[][see Fig.~\ref{fig:ccf-opt-xray}b]{VPV11}. 
Increase of the mass accretion rate obviously causes an increase in the X-ray luminosity, but 
at the same time the  optical synchrotron from the hot flow may drop because of  higher  self-absorption. 
A higher accretion rate can also lead to a decrease of the truncation radius,  
collapse of the hot flow at large radii, and the suppression of the OIR emission.
Both scenarios leads to anti-correlation between optical and X-rays 
and to the negative contribution to the CCF,  with the shape resembling that of the X-ray ACF. 
On the other hand, the  second, reprocessed component correlates with the X-rays, but is delayed and smeared,   
giving rise to a positive CCF peaking at positive lags (optical delay). 
The combined CCF has a complicated shape consistent with the data  (Fig.~\ref{fig:ccf-opt-xray}b). 
The PDS of the optical in this model consists of three components: the synchrotron (which has nearly identical shape to the X-ray PDS), 
the irradiated disc (which has less power at high frequencies because of   smearing) and the cross-term of variable sign. 
The total optical variability is   strongly reduced at low frequencies where the synchrotron and the disc 
vary out of phase  \citep[][see Fig.~\ref{fig:ccf-opt-xray}a]{VPV11}. These PDS shapes are very similar to 
that observed in GX~339$-$4 \citep[see fig.~9 in][]{GDD10}.

Further clues on the origin of the optical emission come from the QPOs seen 
in the light curves of a number of low-mass BHBs 
\citep{Motch83,Motch85,Imam90,SteiCam97,HHC03,DGS09,GDD10}.
The optical, UV and X-ray QPOs in XTE~J1118+480 all have the same frequency, 
which evolves during the two months of observations \citep{HHC03}. 
There is also a clear connection between the optical and X-ray light curves 
in the 2007 data on Swift~J1753.5$-$0127 (see \citealt{DGS08}; A. Veledina et al. in prep.) 
seen as a modulation at the optical QPO frequency in the optical/X-ray CCF. 
Similarly, GX~339$-$4  shows oscillations at the same frequency in the optical and X-rays \citep[][fig.~21]{Motch83,GDD10}. 
The question now arises how it is possible that the X-rays and OIR/UV vary 
at the same frequency and are phase-connected? 
This fact finds a simple explanation in the hybrid hot flow model because  
the hot flow can precess as  a solid body (see Sect.~\ref{sec:xray_variab}) and therefore the long wavelength emission 
produced in the outer part of that flow is related to the X-rays produced 
in the inner part of the same flow \citep{VPI13}.

Recently, periodic eclipses in the optical light curve of BHB Swift~J1357.2$-$093313  were observed \citep{Corral13Sci}. 
If the period is related to the Keplerian frequency of the obscuring region, 
the sharpness of the eclipses implies that the size of the optical emission region is
below 20\,000 km (i.e. $<700 R_{\rm S}$ for a $10\msun$ BH). 
These constraints are easily satisfied in the hot flow scenario.

\subsection{Polarisation}

Polarisation degree and polarisation angle provide two more observational constraints on the emission models. 
The only indication of the X-ray polarisation from BHB goes back to the OSO-8 satellite 
\citep{Weisskopf77}, which measured 3.1$\pm$1.7\% linear polarisation from Cyg X-1 at 2.6~keV.  
Such a polarisation can be produced by Compton scattering if the geometry of the X-ray emitting region is a
flattened disc-like structure \citep[$H/R\sim0.2$ according to the calculations of][]{LS76}, 
consistent with the hot flow scenario.

Recently, strong linear polarisation ($\Pi=67\pm30$\%) 
in the soft  $\gamma$-rays above 400 keV was detected in Cyg X-1 with the IBIS instrument onboard {\it INTEGRAL} \citep{LRW11}.  
Similar polarisation  ($\Pi=76\pm15$\%)  was also observed with the SPI spectrometer \citep{JRCC12}. 
The polarisation angle of $40\degr-42\degr$ is about $60\degr$ away 
from the radio jet axis at $\approx-20\degr$ \citep{JRCC12,ZLS12}.  
Such a large polarisation degree in the MeV range is extremely difficult to get in any scenario. 
Synchrotron jet emission from non-thermal electrons in a highly ordered magnetic field 
can have a large polarisation degree (up to $\sim$70 per cent) in the optically thin part of the spectrum, and indeed 
a high polarisation in the radio and the optical bands reaching 30--50 per cent is observed from extragalactic relativistic jets 
\citep{Impey91,Wills92,Lister01,Marscher02,Ikejiri11}. 
However, this scenario also needs a very hard electron spectrum as well as an extreme fine-tuning to 
reproduce the spectral cutoff at a few MeV  \citep{ZLS12}.  
In the hot-flow scenario, the MeV photons are produced by non-thermal Compton scattering of the  100 keV photons by 
electrons with Lorentz factors $\gamma$$\sim$2--4. 
These electrons cannot be isotropic, because no significant polarisation is expected  in that case \citep{Pou94ApJS}.  
This then implies that they must have nearly one-dimensional motion, e.g. along the  large-scale 
magnetic field lines threading the flow. 
The $60\degr$ offset of the polarisation vector relative to the jet axis then implies the inclined field lines. 
If the measured high polarisation degree is indeed real, 
this would put strong constraints on the physics of particle acceleration in the hot flow 
and the magnetic field geometry. 

In the OIR bands, polarisation is very small and does not exceed  a few per cent  \citep{SHH04,SFW08,RF08,CDR11}. 
In the hot-flow scenario, polarisation degree of the optically thin synchrotron radiation 
in the OIR band is expected to be essentially zero 
(independently of the magnetic field geometry) because the 
Faraday rotation angle  exceeds $10^5$ rad \citep{VPV13}. 
In the optically thick regime, the intrinsic polarisation (parallel to the magnetic field lines) 
is not more than about 10 per cent even for  the ordered magnetic field \citep{PS67,GS69}. 
Faraday rotation in the disc atmosphere can still essentially depolarise that emission.  
The reprocessed emission from the outer disc can be slightly polarised because magnetic field there is smaller. 
If the jet were responsible for the OIR emission,  one would expect instead a much higher polarisation 
because its radiation is optically thin, not consistent with the data.  
The observed small polarisation can also be produced by dust/electron scattering in the source vicinity or by the interstellar dust.

\section{Summary}
\label{sect:summary}

The purely thermal hot-flow model was shown  to be consistent with many X-ray characteristics. 
However, that model fails to account for  the MeV tails and a number of OIR properties. 
Addition of a small, energetically-negligible non-thermal component to the electron distribution dramatically 
changes  the prediction of the model.  
The hybrid hot-flow model is now successful in explaining the following facts: 
\begin{enumerate}
\item stability of spectra with photon index $\Gamma$$\sim$1.6--1.8 and the cutoff at $\sim$100~keV in the hard state 
\citep{PV09,MB09},
\item concave X-ray spectrum \citep{KCG01,VPV13},
\item low level of the X-ray and OIR polarisation \citep{VPV13},
\item presence of the MeV tail in the hard state \citep{PV09,MB09},
\item softening of the X-ray spectrum with decreasing luminosity below $\sim$10$^{-2}L_{\rm Edd}$ \citep{VVP11},
\item weakness of the cold accretion disc component in the hard state, 
\item correlation between the spectral index, the reflection amplitude, the width of the iron line and 
      the frequency of the quasi-periodic oscillations,
\item hard X-ray lags with logarithmic energy dependence \citep{KCG01},
\item non-thermal OIR excesses and flat OIR spectra \citep{VPV13},
\item OIR colours of the flares in the hard state (Poutanen et al., in prep.), 
\item strong correlation between OIR and X-ray emission and a complicated CCF shape \citep{VPV11}, 
\item quasi-periodic oscillations at the same frequency in the X-ray and optical bands \citep{VPI13}.
\end{enumerate}
The model does not explain the radio points and the soft IR spectra. The jet  is obviously a better model for those data. 
We, however, struggle to find any other observational fact that could be in conflict with the hybrid hot flow -- 
truncated cold disc scenario.

Recently, the jet paradigm became popular and it was claimed that the jets are responsible 
not only for the radio emission from the BHBs, but also the OIR and even the X-ray emission. 
Unfortunately, that model is in contradiction with dozens of  observed facts 
\citep[see][and references therein]{VPV13},
which are usually ignored by the model proponents. 
When new data appear, they often are rather puzzling and difficult to understand 
within the available paradigms. However, it would be beneficial for the community when introducing brand new models
to  check also whether those models satisfy other observational constraints.  

In spite of a serious progress in understanding of the viscosity in accretion discs around BHs, 
there are still many open questions. 
If non-thermal particles are present in the hot flow, it is now time to understand 
what is their nature. How are they accelerated: in shocks or in magnetic reconnections events, or maybe 
via diffusive acceleration by MHD turbulence? 
How are they related to the magneto-rotational instability that presumably drives the accretion?  
We hope that  the observational advances will soon be reflected in the advance of the theory.

\begin{acknowledgements}
The work was partially supported by the Academy of Finland grant 268740 (JP)  and 
the Finnish Doctoral Program in Astronomy and Space Physics (AV). 
We thank Tomaso Belloni, Andrzej Zdziarski, and Feng Yuan for valuable comments. 
\end{acknowledgements}


\end{document}